%% file: paper_final.tex
\documentclass[aps,prd,superscriptaddress,floatfix,nofootinbib,preprintnumbers]{revtex4}
\usepackage{amsmath}
\usepackage{graphicx,tabularx}
\usepackage{url}
\usepackage{footmisc}

\renewcommand{\thefootnote}{\fnsymbol{footnote}}

\newcommand{\ie}{{\sl i.e. }}
\newcommand{\eg}{{\sl e.g. }}

\newcommand{\qss}{\left\langle {\cal{Q}}_{ {\cal{S}} + {\cal{B}} } \right\rangle}

\newcommand{\qsb}{ {\cal{Q}}_{ {\cal{S}} + {\cal{B}} }}
\newcommand{\qbb}{ {\cal{Q}}_{\cal{B}} }

\include{declare}

\begin{document}

\title{Jets plus Missing Energy with an Autofocus}

\author{Christoph Englert}
\affiliation{Institut f\"ur Theoretische Physik, Universit\"at Heidelberg, Germany}

\author{Tilman Plehn}
\affiliation{Institut f\"ur Theoretische Physik, Universit\"at Heidelberg, Germany}

\author{Peter Schichtel}
\affiliation{Institut f\"ur Theoretische Physik, Universit\"at Heidelberg, Germany}

\author{Steffen Schumann}
\affiliation{Institut f\"ur Theoretische Physik, Universit\"at Heidelberg, Germany}

\begin{abstract}
  Jets plus missing transverse energy is one of the main search
  channels for new physics at the LHC. A major limitation lies in our
  understanding of QCD backgrounds. Using jet merging we can describe
  the number of jets in typical background channels in terms of a
  staircase scaling, including theory uncertainties. The scaling
  parameter depends on the particles in the final state and on cuts
  applied.  Measuring the staircase scaling will allow us to also
  predict the effective mass for Standard Model backgrounds. Based on
  both observables we propose an analysis strategy avoiding model
  specific cuts which returns information about the color charge and
  the mass scale of the underlying new physics.
\end{abstract}

\maketitle

\section{Jets with missing energy}
\label{sec:inclsearch}

Missing transverse energy is a general signature for dark matter
related new physics at hadron colliders~\cite{review}. It has a long
history at the Tevatron and to date gives the strongest bounds on
squark and gluino masses in supersymmetric extensions of the Standard
Model. At the LHC the first new exclusion limits for squarks and
gluinos have recently appeared, in the CMSSM toy model as well as in a
more general setup~\cite{atlas,cms,jay}.  All of these analyses are
based on jets plus missing energy including a lepton veto which
constitutes the most generic search strategy for strongly interacting
new particles decaying into a weakly or super-weakly interacting dark
matter candidate~\cite{jetetmiss,review}.

While the first results are based on very inclusive cuts, following
the ATLAS~\cite{cscnotes} and CMS~\cite{cmstdr} documentations we
expect more specific analyses to appear soon. The reason is that in
their current form the analyses can and should be optimized for
specific new physics mass spectra. More specialized analyses for jets
plus missing energy rely on a missing transverse momentum cut and on a
certain number of staggered jet transverse momentum
cuts~\cite{cscnotes,cmstdr}. Unfortunately, they are therefore hard to
adapt to modified mass spectra and by definition show a poor
performance for not optimized model parameters. In addition, they are
counting experiments in certain phase-space regions, which means that
for any additional information on the physics behind an anomaly we
have to wait for a dedicated analysis.\bigskip

A major problem of searches for new physics in pure QCD plus missing
energy final states is the prediction of background
distributions. Aside from the improved signal-to-background ratio this
is one of the reasons why applying fairly restrictive cuts on the
number of jets and on their transverse momentum is a promising
strategy. Such cuts relieve us from having to understand the
complete $p_T$ spectra~\cite{early_matching} of general exclusive or
inclusive $\nj$-jet events at the LHC. Experimentally, however, we
should by now be in a position to simulate these distributions using
the {\sc Ckkw}~\cite{ckkw,mets_matching} or {\sc Mlm}~\cite{mlm}
matching methods implemented in {\sc Sherpa}~\cite{sherpa}, {\sc
  Alpgen}~\cite{alpgen}, or {\sc MadEvent}~\cite{madgraph}. The
different approaches have been compared in some detail, for example
for $W$+jets production~\cite{mcreview,matching_comparison}.  What is
still missing is a systematic study of theory uncertainties in
multi-jet background simulations for top quark analyses and new
physics searches, \ie including large jet multiplicities down to
intermediate jet transverse momenta, but reflecting a well defined
hard scale given by the signal process. Motivated by theoretical and
statistical considerations we define all observables as exclusive,
specifically in the number of jets.

In this paper we establish a proper simulation of multi-jet processes
and estimate these theory uncertainties, with a focus on the question
what actually constitutes the theory error. This way LHC data in
control regions can be used to understand very generic scaling
features ({\sl staircase scaling}\footnote{Staircase scaling for jet
  rates is often referred to as Berends scaling. However, to our best
  knowledge it was first introduced and discussed by the authors of
  Ref.~\cite{scaling1}.}) which have already been observed in
data~\cite{scaling_tev,wjets} and which we can extend based on
appropriate Monte Carlo studies. This staircase scaling we can
reproduce and study using QCD Monte Carlo simulations, including
different hard processes and the effects of cuts. Combining these
simulations with LHC data should give us a quantitative handle on
multi-jet rates in many applications.

Moreover, we can use our knowledge about the exclusive $\nj$
distributions to predict other notoriously difficult multi-jet
observables.  So once we understand the uncertainties on the multi-jet
spectra we turn to the effective mass. In its many incarnations it
either includes the leading jet or it does not and is either limited
to four jets or any other number of jets~\cite{alan_chris}. Obviously,
any specific definition of this mass variable increases its
sensitivity to theory uncertainties. We study the most generic
definition of the effective mass including {\sl all jets} visible
above a transverse momentum threshold. The theory uncertainties of
this observable can be closely linked to the jet-multiplicity
distribution.  Using the scaling properties of the exclusive jet
multiplicities we can strongly reduce the theory uncertainty in the
effective-mass spectrum in a consistent manner. The same should be
true for other variables which we can use to extract new physics from
jet dominated backgrounds.

Similar questions are currently being asked to control regions in a
purely data-driven approach. However, the conversion from background
regions into the signal region either by shifting the kinematic regime
or by changing the hard core processes off which we radiate jets
requires a good understanding for example of the effects of background
rejection cuts and of background sculpting features in the definition
of these observables. These effects we can reliably estimate in an
appropriate Monte Carlo study and then combine for example with an
over-all normalization from data.\bigskip

Finally, we suggest an analysis strategy which on the one hand makes
maximum use of the jet patterns and on the other hand does not require
any tuning of cuts. The only ingredient of our analysis which does not
involve jets is a missing transverse energy cut to reduce pure QCD
backgrounds and an isolated lepton veto against $W$+jets
backgrounds. To reduce both of them to a manageable level we require
\begin{equation}
\met > 100~\gev \qquad \qquad
\text{and a lepton veto if}\quad 
p_{T,\ell} > 20~\gev,\;|y_\ell| < 2.5
\label{eq:met}
\end{equation}
as the basic and only electroweak cuts to reduce the QCD background.
The exact numbers are not very dependent on the details of the model 
as long as the new physics sector provides a WIMP dark matter candidate. 
To account for fake missing energy from QCD jets we apply an additional
factor of $1/500$ for pure QCD and hadronic top-quark final
states. This rough fake rate we estimate from Ref.~\cite{atlas}. It
provides us with a rather conservative estimate compared \eg to
Ref.~\cite{cmstdr}.

After these very generic acceptance cuts a two-dimensional correlation
of the effective mass vs the exclusive jet multiplicity is the
appropriate distribution to extract limits on strongly interacting new
physics or in the case of an excess study the mass scale as well as
the color charge of the new states. Because all our observables are
defined jet-exclusively we can to a good approximation study this
two-dimensional distribution using a log-likelihood shape
analysis. The contributions of different regions in the $\nj$-$\meff$
space to the binned log-likelihood automatically focus on the correct
phase-space region and are readily available for improved analyses as
well as theoretical interpretation.

\section{Jet number scaling}
\label{sec:jetcount}

To separate new physics events from a QCD sample after some very basic
cuts we have to understand the number of jets and their energy or
$p_T$ spectra.  This will allow us to exploit two distributions: the
number of jets observed ($\nj$) and their effective mass ($\meff$),
where the definition of the latter usually requires us to define the
number and hardness of the jets included in its construction. Our
maximally inclusive approach means that aside from the fiducial volume
of the detectors all we fix is the algorithmic jet definition to count
a jet towards each of the two measurements. Throughout this paper we
define jets using the anti-$k_T$ algorithm~\cite{antikt} in {\sc
  FastJet}~\cite{fastjet} with a resolution $R_{\text{anti}-k_T}=0.4$
and then require
\begin{equation}
p_{T,j} > \ptmin = 50~\gev
\qquad \qquad 
\text{and}
\qquad\qquad 
|y_j| < 4.5 \; .
\label{eq:jetcuts}
\end{equation}
This defines which jets are counted towards $\nj$ as well as the
$\meff$ distribution.  Given $\ptmin$ we can then evaluate the
2-dimensional $\nj$ vs $\meff$ plane in Section~\ref{sec:autofocus}
using a binned log-likelihood approach.\bigskip

Before we can use the $\nj$ distribution to extract new physics in
the jets plus missing energy sample at the LHC we need to show
that we understand this distribution in detail. Obviously, the
overall normalization of this distribution is not critical. For any 
kind of new physics not completely ruled out by the Tevatron experiments 
the two jet and three jet bins are practically signal free. So the
question becomes: what can we say about the shape of $d\sigma/d\nj$.

For $W$+jets events this kind of distribution has been studied, even
at the LHC~\cite{wjets}. We observe the {\sl staircase
  scaling}~\cite{scaling1,scaling2}, an exponential drop in the
inclusive $\nj$ rates with constant ratios
$\hat{\sigma}_{n+1}/\hat{\sigma}_n$. The numerical value of this ratio
is obviously strongly dependent on $\ptmin$. The original staircase
scaling describes inclusive jet rates, \ie it uses $\hat{\sigma}_n$
including all events with at least $n$ jets fulfilling
Eq.(\ref{eq:jetcuts}). In the light of recent advances in QCD and
because our likelihood analysis should be based on independent bins we
define the scaling in terms of exclusive jet rates, \ie counting only
events with exactly $n$ jets fulfilling Eq.(\ref{eq:jetcuts}) towards
$\sigma_n$. This preserves the normalization of the $\nj$ histogram as
$\sigma_\text{tot} = \sum_n \sigma_n$ and makes it possible to add the
bins in the computation of the log-likelihood. It is interesting to note
that staircase scaling defined either way implies staircase scaling
using the other definition, and that the jet-production ratios of the
two approaches are identical. If we define the universal exclusive
staircase-scaling factor as
\begin{equation}
R \equiv R_{(n+1)/n} = \frac{\sigma_{n+1}}{\sigma_n} \; ,
\label{eq:def_r}
\end{equation}
we find for the usual inclusive scaling denoted by a hat over all parameters
\begin{alignat}{5}
\hat{R} \equiv \dfrac{\hat{\sigma}_{n+1}}{\hat{\sigma}_n} 
&= \dfrac{\sigma_{n+1} \sum_{j=0}^\infty R^j}{\sigma_n + \sigma_{n+1} \sum_{j=0}^\infty R^j} 
&= \dfrac{R \sigma_n}{(1-R) \sigma_n + R \sigma_n} 
 = R \; .
\end{alignat}
The same relation we find when we include a finite upper limit to the
number of jets in the sum over $j$. Note, however, that this argument
only holds for a strict staircase scaling where the ratio
$R_{(n+1)/n}$ does not depend on the number of jets $n$. For our
analysis this means that we can use the staircase scaling for a
statistical analysis of the $d\sigma/d\nj$ distribution either in its
inclusive or in its exclusive version, the latter based on independent
$\nj$ bins.\bigskip

In this section we will show that (1) such a scaling exists not only
for $W/Z$+jets but also for pure QCD events and (2) we can reliably
estimate the scaling factor and possible deviations from it from
theory. A purely data-driven background analysis of this distribution
might be possible and should be combined with our results. For
example, we can one by one remove the missing energy cut and the
lepton veto in Eq.(\ref{eq:met}) which gives us background dominated
event samples to a reasonably large number of hard jets. Adding the
background rejection cuts will then have an impact on the scaling,
which we can estimate reliably. For the signal hypothesis we have to
entirely rely on QCD predictions.\bigskip

\begin{figure}[t!]
\includegraphics[width=0.49\textwidth]{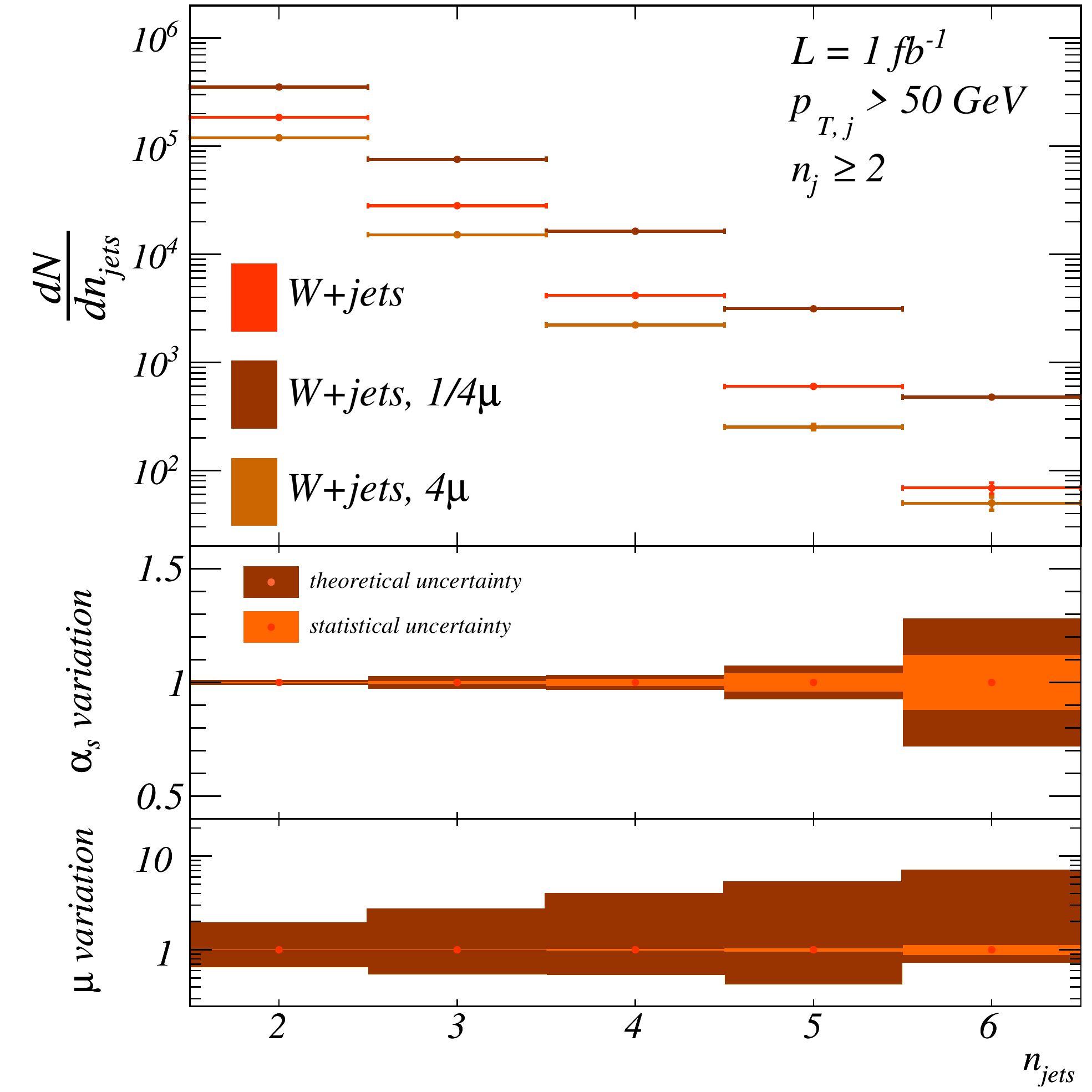}
\includegraphics[width=0.49\textwidth]{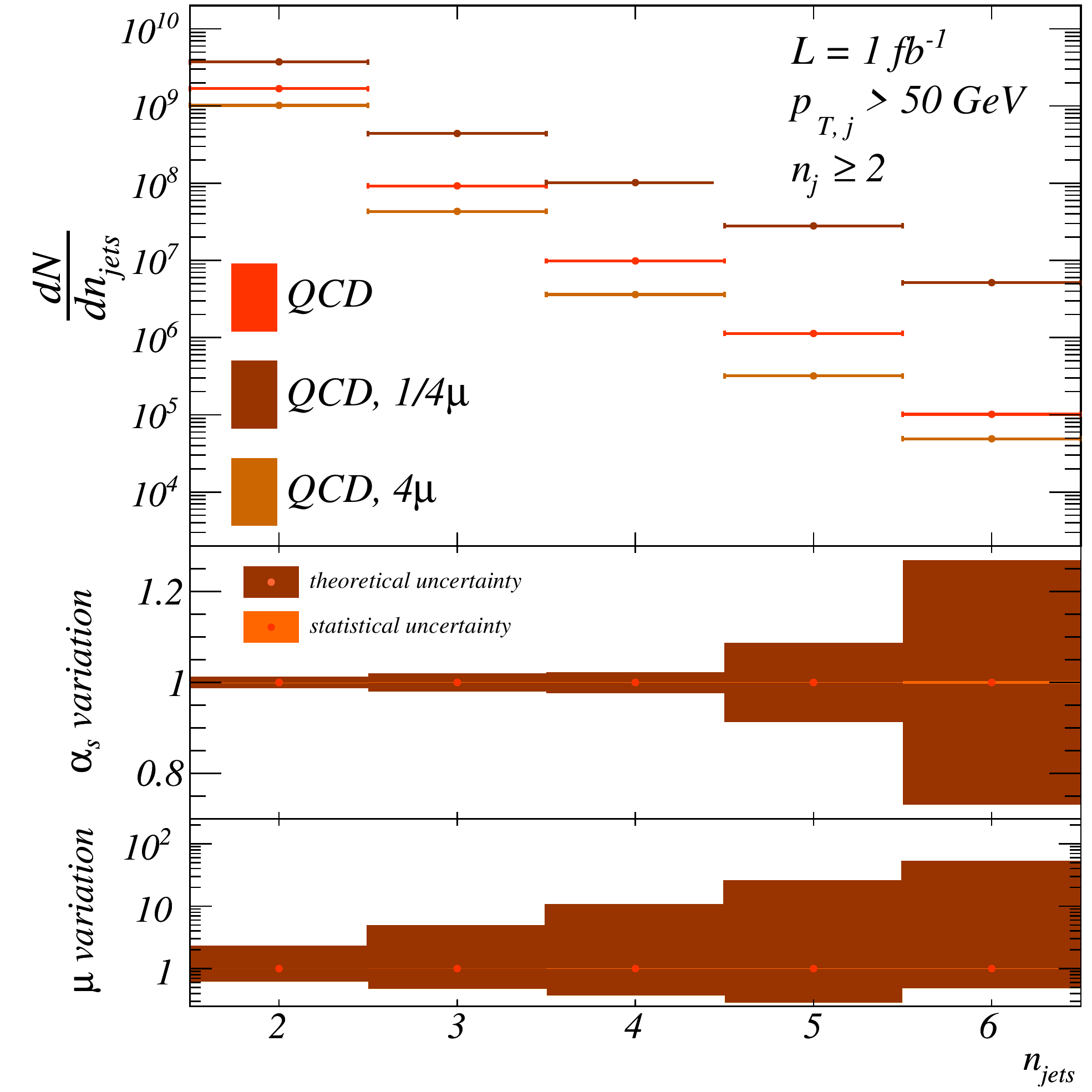}
\caption{\label{fig:scale_light} Exclusive $d\sigma/d\nj$ distribution 
  for $W$+jets (left) and QCD jets production at the LHC. Only the jet 
  cuts given in Eq.(\ref{eq:jetcuts}) are applied, neither a $\met$ 
  cut nor a lepton veto is imposed. The second panel shows the parametric 
  uncertainty due to a consistent change of $\alpha_s(m_Z)$ between 0.114 
  and 0.122. The third panel shows the reach of a consistent scale factor 
  treatment which can be experimentally determined and should not be 
  considered a theory uncertainty.}
\end{figure}

As a starting point we discuss the established staircase scaling in
$W$+jets production. The behavior of $Z$+jets is exactly the same. For
our analysis we produce {\sc Ckkw}-matched~\cite{mets_matching}
background samples for $W$+jets (to 5 ME jets), $Z$+jets (to 5 ME
jets), $t\bar t$+jets (to 2 ME jets), and QCD jets (to 6 ME jets) with
{\sc Sherpa}-1.2.3~\cite{sherpa}.
Higher order corrections to the
inclusive scaling we expect to, if anything, improve the assumption of
a constant jet ratio $\hat{R}$ for example in $W$+jets
production~\cite{blackhat}.

In the left panel of Figure~\ref{fig:scale_light} we show the
exclusive $\nj$ distribution for the LHC running at 7~TeV center of
mass energy. To increase our statistics to large enough values of
$\nj$ we do not apply the selection cuts Eq.(\ref{eq:met}) in this
first step. We already see that we can qualitatively fit a line
through the central points on a logarithmic axis for each set of input
parameters.

Before we quantitatively evaluate this scaling we need to consider the
uncertainties associated with our simulation.  This is crucial if we
want to use the $\nj$ scaling as a background estimate for new physics
searches in QCD final states. There are two distinct sources of
uncertainty in our simulation.  First, there exists a parametric
uncertainty, namely the input value of $\alpha_s(m_Z)$ or some other
reference scale. To address this, we consistently evaluate the parton
densities around the central NLO value $\alpha_s(m_Z) = 0.118$ inside
a window $0.114 - 0.122$~\cite{cteq10} and keep this value for all
other appearances of the strong coupling in our matrix-element plus
parton-shower Monte-Carlo simulations. In Figure~\ref{fig:scale_light}
we see that the resulting error bar on the $d\sigma/d\nj$ increases
with the number of jets, but stays below 30\% even for the radiation
of six jets.  For luminosities around $1~\ifb$ the error on $\alpha_s$
is roughly of the same order as the experimental statistical
error. Systematic errors we do not consider, even though they will at
some point dominate over the statistical errors. After any kind of
realistic background rejection the combined experimental error will
exceed the parametric $\alpha_s$ uncertainty.\bigskip

The reason why we cannot use staircase scaling in $W$+jets to measure
$\alpha_s$ is a second source of QCD uncertainty: aside from the
parametric $\alpha_s$ error band, an actually free parameter in our
QCD simulation is a common scaling factor $\mu/\mu_0$ in all
appearances of the factorization and renormalization scales, including
the starting scale of the parton shower. Identifying all scales
follows the experimental extraction of the parton densities and
$\alpha_s$ in a simultaneous fit. The interpretation of DGLAP
splitting in terms of large logarithms tells us that the factorization
and renormalization scales have to be identified with the transverse
momentum of the radiated jets. By definition, such leading-logarithm
considerations leave open the proportionality factor in the relation
$\mu \propto |p_{T,j}|$. Any constant factor can be separated from the
dangerous logarithm as a non-leading constant value.

Because this constant cannot be derived from first principles we vary
it in the range $\mu/\mu_0 = 1/4 - 4$ and show the numerical result in
Figure~\ref{fig:scale_light}. As expected, the variation of the jet
rates with this scaling parameter is huge --- much larger than the
experimental uncertainties we expect from the LHC and which we know
from the Tevatron.  In Figure~\ref{fig:scale_light} we can first of
all check that introducing such a scaling factor does not seriously
impact the observed staircase scaling. Counting such a constant
towards the theory uncertainty is questionable if we can determine it
experimentally. For example for {\sc Sherpa} we know from Tevatron
that the scaling factor should essentially be
unity~\cite{sh_validation}, which in the spirit of Monte-Carlo tuning
means that for example in {\sc Sherpa} the naive default parameter
choice comes out as correctly describing the data. Of course, this
does not have to be true for other simulation tools.  An interesting
question to ask once we have access to it at the LHC would be if this
{\sl per se} free parameter really is the same for different channels,
like $W/Z$+jets and QCD jets.\bigskip

In the right panel of Figure~\ref{fig:scale_light} we show the same
distributions for pure QCD jet production. Again, not applying the
cuts in Eq.(\ref{eq:met}) we observe staircase scaling, however, with
some caveats for the two and three jet bins. This is related to the
definition of the hard process.  As expected, the scale factor
$\mu/\mu_0$ has very large impact not on the existence of a staircase
scaling but on the jet ratio $R$. The parametric uncertainty due to
the error bar on $\alpha_s(m_Z)$ is again small once we vary the
strong coupling consistently everywhere, staying below 30\% for up to
six jets. The parametric uncertainty for the pure QCD case and the
$W$+jets case is clearly very similar. The scale factor variation
$\mu/\mu_0 = 1/4 - 4$ gives an even larger band of possible ratios of
cross sections, to be contrasted with a reduced statistical uncertainty
compared to the $W$+jets case. Our argument that this over-all scale
factor should be determined experimentally is therefore even more
applicable for the QCD case. To date such an analysis does not exist, so
while in the following we will use unity as the appropriate scale
factor for {\sc Sherpa} this needs to be verified
experimentally.\bigskip

\begin{table}[t]
\begin{tabular}{ll|rrrrrr|rr}
\hline
channel (cuts) &
& \rule[0cm]{0cm}{4ex} 
$R_{2/1}$ & $R_{3/2}$ & $R_{4/3}$ & $R_{5/4}$ & $R_{6/5}$ & $R_{7/6}$ & 
$R_0$ & $\dfrac{dR}{d\nj}$ \\[4mm]
&& \multicolumn{6}{c|}{{\sc Sherpa} simulation} 
&  \multicolumn{2}{c}{linear fit} \\
\hline
$W$+jets ($p_{T,j}>50~\gev$) \rule[0cm]{0cm}{3ex} 
&  & $0.1931(3)$ & $0.1494(5)$ & $0.157(1)$ & $0.138(3)$ & $0.115(8)$ & $0.09(2)$ & $0.150(1)$ & $-0.001(1)$ \\ 
$W$+jets ($+$~lepton veto) \rule[0cm]{0cm}{3ex} 
&  & $0.2290(4)$ & $0.1494(7)$ & $0.164(2)$ & $0.139(4)$ & $0.12(1)$ & $0.09(2)$ & $0.149(1)$ & $-0.002(1)$ \\ 
$W$+jets ($+~\met>100~\gev$) \rule[0cm]{0cm}{3ex} 
&  & $0.252(1)$ & $0.224(2)$ & $0.190(5)$ & $0.16(1)$ & $0.15(2)$ & $0.09(4)$ & $0.239(3)$ & $-0.032(3)$ \\ 
\hline
$Z$+jets ($p_{T,j}>50~\gev$) \rule[0cm]{0cm}{3ex} 
&  & $0.1463(2)$ & $0.1504(6)$ & $0.147(1)$ & $0.138(4)$ & $0.123(9)$ & $0.07(2)$ & $0.154(1)$ & $-0.006(1)$ \\ 
$Z$+jets ($+~\met > 100~\gev$) \rule[0cm]{0cm}{3ex} 
&  & $0.2251(6)$ & $0.185(1)$ & $0.166(3)$ & $0.154(6)$ & $0.14(1)$ & $0.08(3)$ & $0.193(2)$ & $-0.018(2)$ \\
\hline
QCD jets ($p_{T,j}>50~\gev$) \rule[0cm]{0cm}{3ex}  
&   & --- & $0.0552(1)$ & $0.1074(5)$ & $0.106(1)$ & $0.125(5)$ & $0.12(1)$ & $0.105(2)$ & $0.001(1)$\\
\hline
$(t \bar{t})_{hh}$+jets ($p_{T,j}>50~\gev$)  \rule[0cm]{0cm}{3ex} 
&   & $3.69(9)$ & $1.26(2)$ & $0.67(1)$ & $0.366(9)$ & $0.24(1)$ & $0.15(5)$ &\\ 
\hline
$(t \bar{t})_{\ell \ell/h}$+jets ($p_{T,j}>50~\gev$) \rule[0cm]{0cm}{3ex} 
&  & $1.96(2)$ & $0.851(7)$ & $0.465(5)$ & $0.260(5)$ & $0.168(8)$ & $0.12(2)$ &\\ 
$(t \bar{t})_{\ell \ell/h}$+jets ($+$~lepton veto) \rule[0cm]{0cm}{3ex} 
&  & $1.75(2)$ & $0.765(10)$ & $0.391(7)$ & $0.228(8)$ & $0.14(1)$ & $0.12(3)$ &\\ 
$(t \bar{t})_{\ell \ell/h}$+jets ($+~\met>100~\gev$) \rule[0cm]{0cm}{3ex} 
&  & $1.60(5)$ & $0.83(2)$ & $0.49(2)$ & $0.25(2)$ & $0.15(2)$ & $0.19(7)$ &\\ 
\hline
\end{tabular}
\caption{Jet ratios for all Standard Model channels, including
  (semi-)leptonic and hadronic top pairs for the central scale 
  choice $\mu=\mu_0$. The quoted errors are statistical errors from the 
  Monte Carlo simulation. The numbers correspond to the curves shown in
  Figures~\ref{fig:scale_light} and~\ref{fig:njetfits_Z}.}
\label{tab:ratios}
\end{table}

\begin{figure}[b!]
\includegraphics[width=0.32\textwidth]{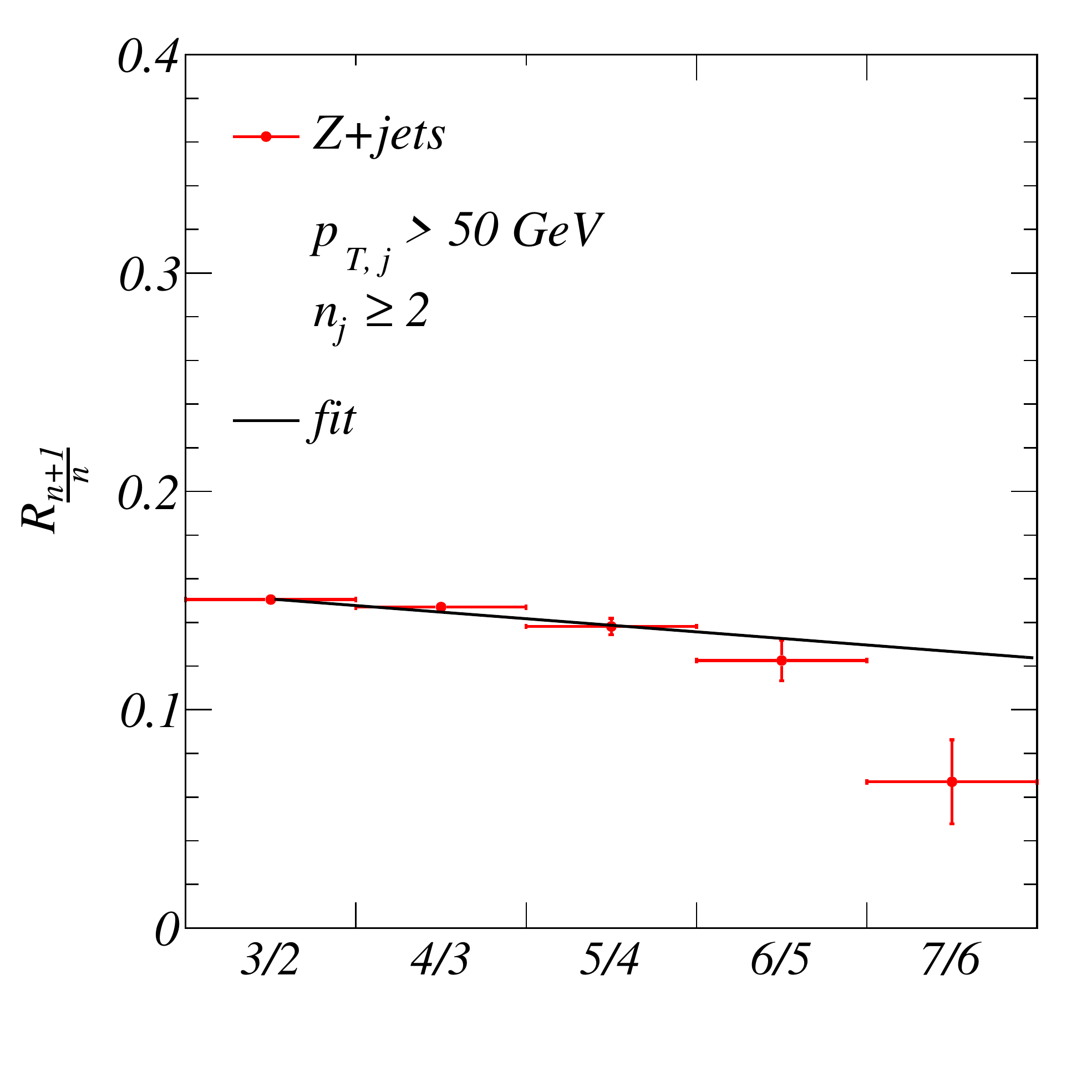}
\includegraphics[width=0.32\textwidth]{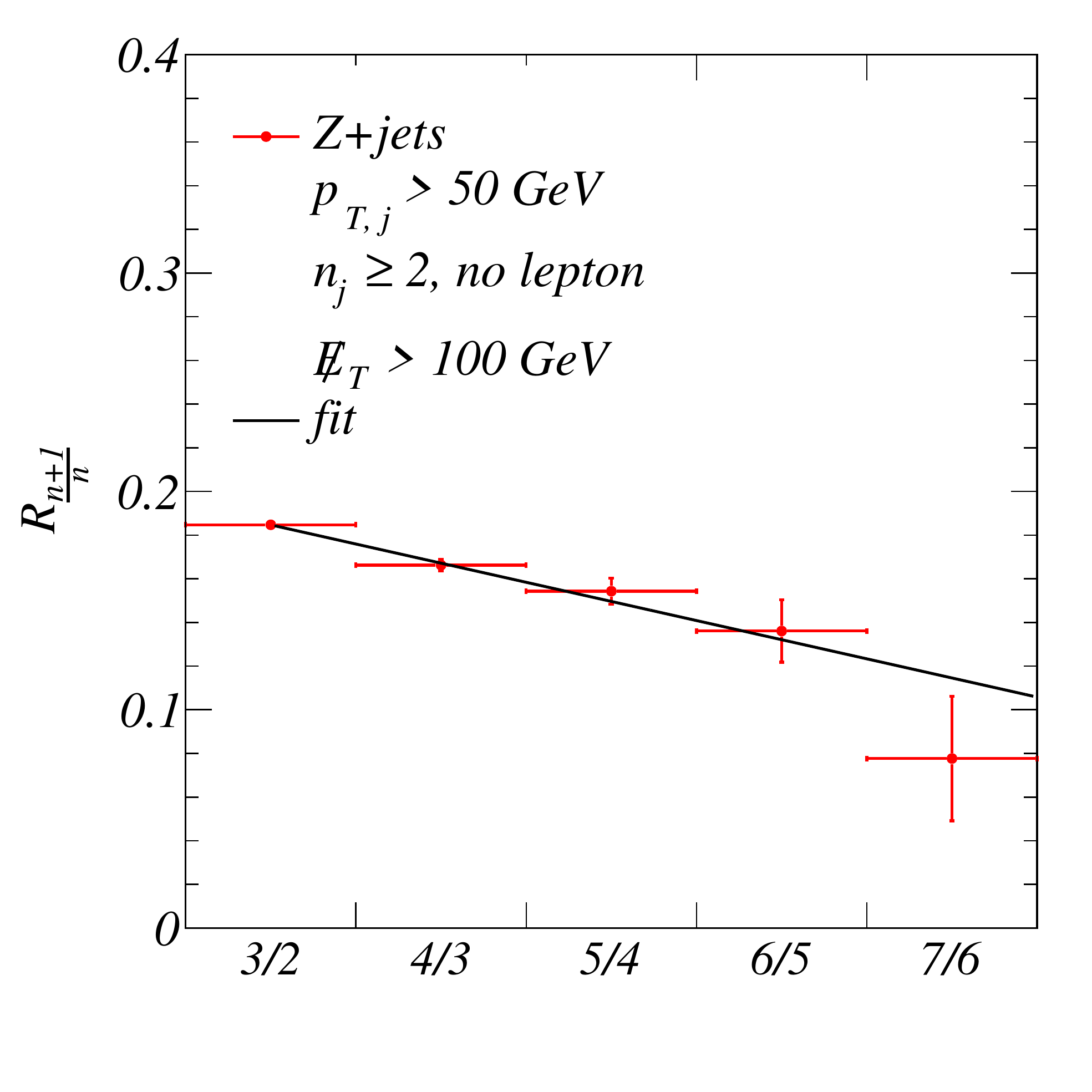}
\includegraphics[width=0.32\textwidth]{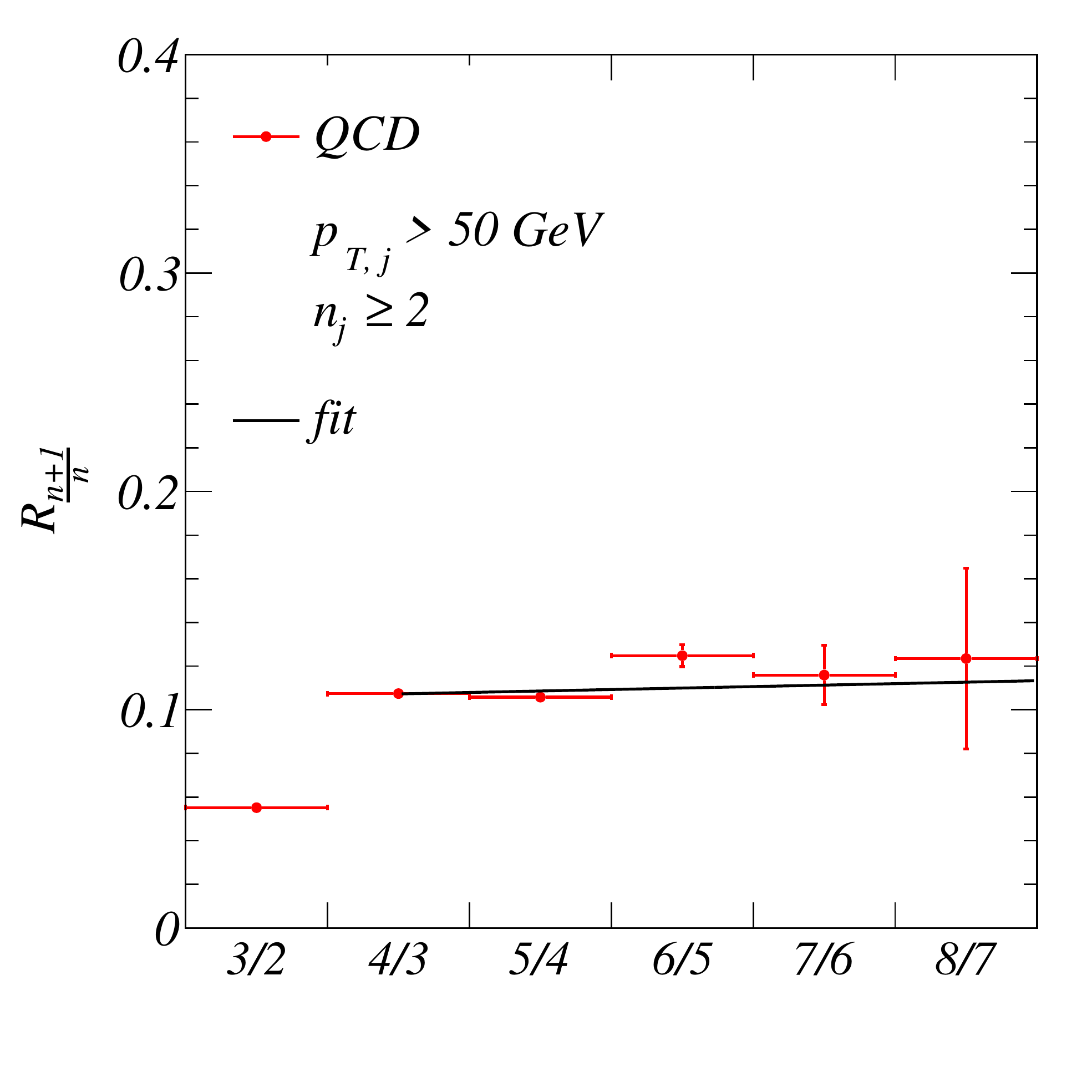}
\caption{Jet ratios for $Z$+jets production (without and with the
  $\met > 100$~GeV cut) and QCD jets production, corresponding to the
  numbers listed in Table~\ref{tab:ratios}. The error bars indicate
  the remaining Monte Carlo uncertainty in our simulation.}
\label{fig:njetfits_Z} 
\end{figure}

Once we understand the size of theory uncertainties for the exclusive
$d \sigma/d \nj$ distribution we need to quantify the quality of the
observed staircase scaling. Since the quantitative outcome will depend
on the background rejection cuts we apply, we study the scaling
without the cuts shown in Eq.(\ref{eq:met}), after the lepton veto
only, and including the lepton veto as well as the missing energy cut.
Starting from the individual $R_{(n+1)/n}$ values we fit a line
through all relevant data points, as a function of $\nj$
\begin{equation}
R(\nj) = R_0 + \frac{dR}{d\nj} \; \nj \; ,
\label{eq:fit}
\end{equation}
and determine the slope to compare it to our prediction $dR/d\nj
=0$.\bigskip

 In Table~\ref{tab:ratios} we list the exclusive jet ratios as
shown in Figure~\ref{fig:scale_light}. For the $W/Z$+jets case we see
that the radiation of one compared to a second jet off the Drell-Yan
process $R_{2/1}$ does not show this scaling. The reason for this
specific feature is the definition of the hard core process alluded to
before. To generate the relatively hard jets and the
large missing energy mimicking the signal events we need to at least
consider $W/Z$+1~jet as the core process.  In addition, we do not take
into account any separation criterion between the first jet and the
gauge boson, which means we treat $\sigma_1$ different from all other
$\sigma_n$.  In Table~\ref{tab:ratios} we see that we are lucky for
the $Z$+jets case, but we are not for the $W$+jets case. The tricky
definition of the hard process $\sigma_1$ as the base of additional
jet radiation suggests that we start our staircase scaling analysis
with $R_{3/2}$.

The statistical uncertainties which we show in Table~\ref{tab:ratios}
and which enter the fit of the slope as defined in Eq.(\ref{eq:fit})
always increase towards larger jet multiplicity. This is an effect of
the way we simulate these events which completely corresponds to an
experimental analysis. If we generate (or measure) all $\nj$ bins in
parallel the first bin will always have by far the smallest
error. Therefore, it determines the constant scaling factor $R_0$ in
our fit as well as in a measurement. For larger values of $\nj$ we
become statistics dominated, which means that Monte Carlo simulations
can extend the reach of actual measurements at any given point in time
towards larger jet multiplicities. This is the phase space region in
which we need to provide new physics searches at the LHC with accurate
background estimates.

Some of the rows listed in Table~\ref{tab:ratios} we also depict in
Figure~\ref{fig:njetfits_Z}. For electroweak gauge boson production we
see that without any cuts $W$ and $Z$ production show the same scaling
parameter $R_0$ as well as a small negative slope.  Within errors the
staircase scaling holds to six and possibly seven jets, even though we
see a slight slope developing towards larger numbers of jets. This is
a phase space effect which is expected once we start probing gluon
parton densities and their sharp drop towards larger parton momentum
fractions and which is well modeled by our computation.

Adding the lepton veto does not change the staircase scaling at
all. This means that forcing the $W$ boson to decay into one fairly
soft lepton and a harder neutrino does not affect the behavior of the
recoiling jets. Adding a significant $\met$ cut, on the other hand,
has a measurable effect on the jet ratios as well as on the slope. For
experimental applications of this scaling, however, it is important to
note that the phase space effects for large $\nj$ as well as the
effect of kinematic cuts are completely described by our
simulation.\bigskip

For pure QCD events we find a remarkable agreement with the staircase
scaling hypothesis, which seems to be supported by recent LHC
analyses~\cite{atlas_jets}. The definition of the hard core process is
somewhat problematic since there exists no inherent hard scale in the
$2\to 2$ process and the infrared behavior of $s$-channel and
$t$-channel diagrams is very different. Therefore, we define
$\sigma_3$ as the starting point of our analysis and $R_{4/3}$ as the
first relevant cross section ratio. Table~\ref{tab:ratios} and
Figure~\ref{fig:njetfits_Z} show that the ratios $R_{(n+1)/n}$ are
essentially constant to eight jets. The slope within statistical
uncertainties is, in contrast to $W/Z$ production, fully compatible
with zero. The central $R_0$ values for $W/Z$+jets and QCD jets
production are slightly different, which is expected by the different
core processes and by the different background rejection cuts.

\section{Decay jets vs jet radiation}
\label{sec:heavy}

In contrast to the QCD and gauge-boson background $\nj$ distributions
from heavy particles decaying to jets include two sources of jets:
first, there are decay jets, which dependent on the spectrum might or
might not be hard enough to stick out. Second, there is QCD jet
radiation, which for heavy states will generically be relatively hard
and dominated by collinear splitting in the initial
state~\cite{skands,madevent_susy}, leading to a non-zero maximum value
of the number of expected initial-state radiation
jets~\cite{review,sgluons}. Due to the hard scale of new-physics
processes on the one hand and because we need to simulate
supersymmetric decays inclusively we best generate the new-physics
events with {\sc Herwig}++-v2.4.2~\cite{herwigpp} and normalize the
cross sections with {\sc Prospino2.1}~\cite{prospino}. All
supersymmetric mass spectra we generate with {\sc
  SoftSusy}~\cite{softsusy} using the SLHA output format \cite{SLHA}
and use {\sc Sdecay}~\cite{sdecay} to calculate the leading-order
branching ratios. We check the jet-radiation results from the {\sc
  Herwig}++ shower with {\sc Mlm} merging implemented in {\sc
  MadEvent}~\cite{madevent_susy}, using {\sc Pythia}~\cite{pythia} for
parton showering and hadronization. As expected, the two simulations
agree well within their uncertainties.\bigskip

\begin{figure}[b!]
\includegraphics[width=0.32\textwidth]{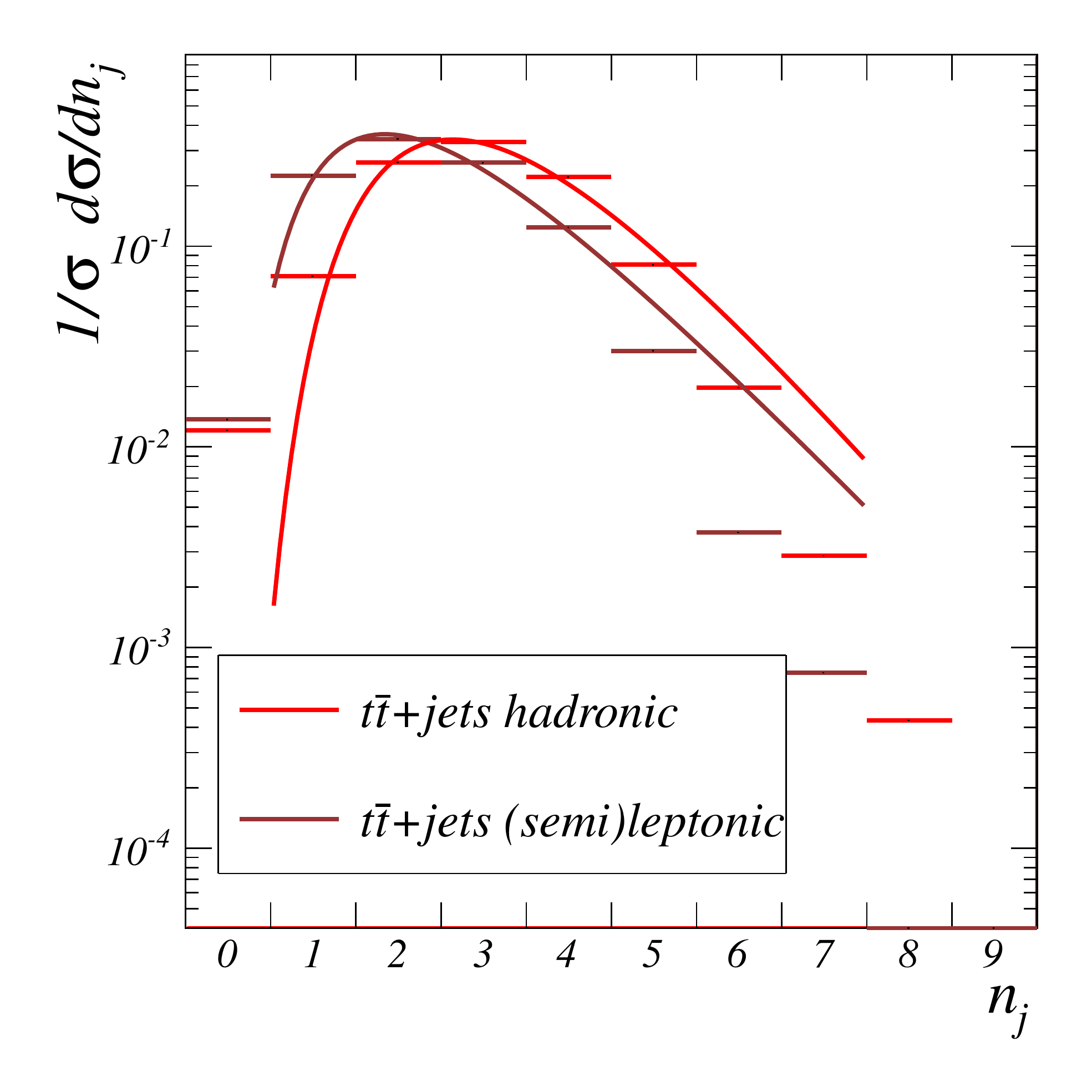}
\includegraphics[width=0.32\textwidth]{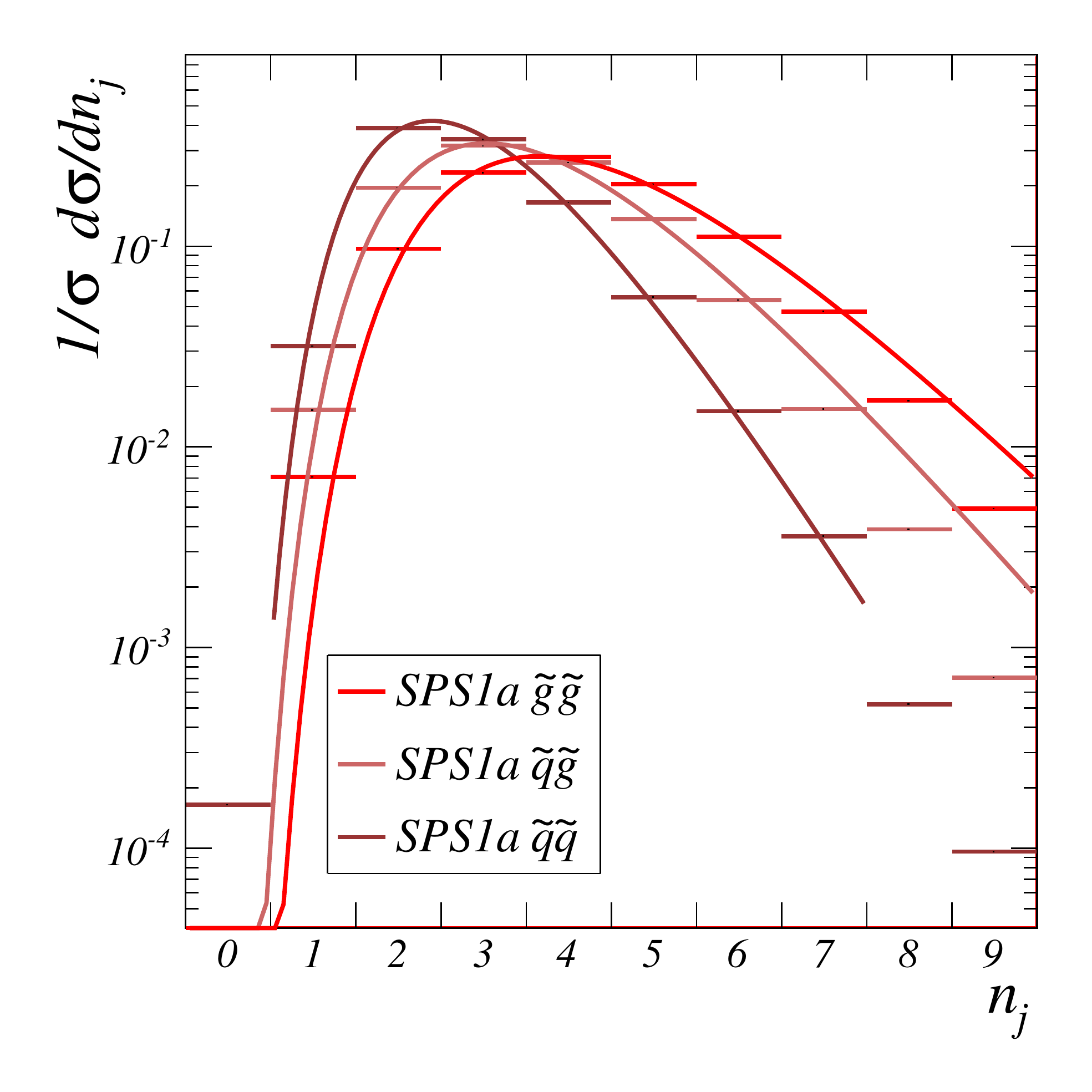}
\includegraphics[width=0.32\textwidth]{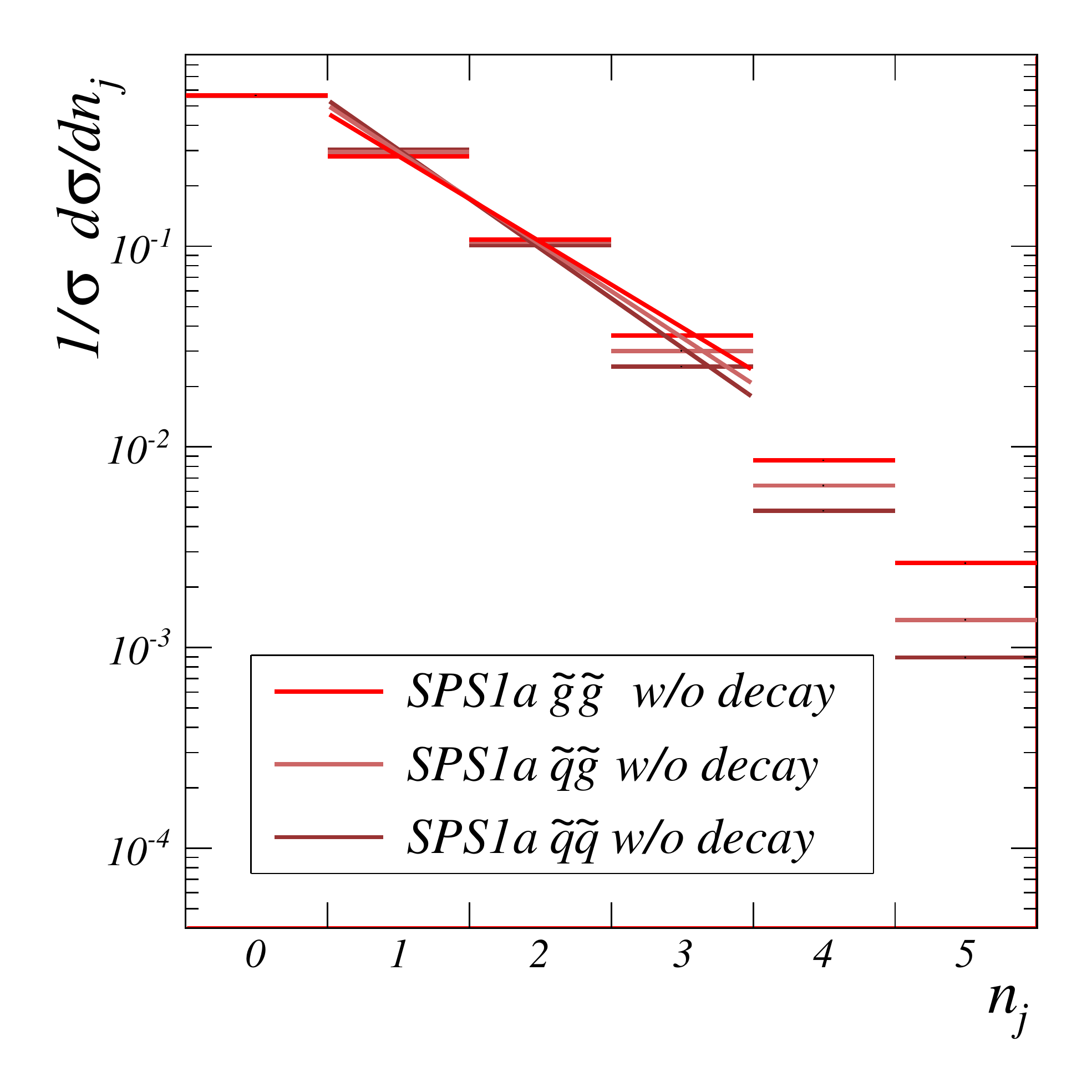}
\caption{Normalized exclusive $d\sigma/d\nj$ distributions for top
  pairs (left) and supersymmetric particle production. For the latter
  we show all decay jets plus QCD jet radiation (center) as well as
  QCD jet radiation only (right). Jets are counted once they fulfill
  Eq.(\ref{eq:jetcuts}).}
\label{fig:scale_heavy} 
\end{figure}

The question for heavy-particle production is how universal its $\nj$
distributions are when we consider Standard Model as well as new-physics 
particles with different masses and color charges, like top
quarks, squarks and gluinos. In Figure~\ref{fig:scale_heavy} we first
show the $\nj$ distributions for (semi-)leptonic and hadronic top-pair
production. We see how all unsubtracted distributions show maxima away
from $\nj = 0$, driven by the presence of decay jets plus relatively
hard jet radiation. In addition, they do not show a staircase scaling
at large jet multiplicities.  Because the particles produced in the
hard process have non-negligible masses even compared to the hadronic
center-of-mass energy the phase-space suppression for example due to
rapidly dropping gluon densities kicks in immediately and bends the
otherwise exponential fall-off.

In the Standard Model we can fit the (semi-)leptonic and purely
hadronic top-pair distributions for all jets fulfilling
Eq.(\ref{eq:jetcuts}) to the function
\begin{equation}
 \frac{d \log \sigma(\nj)}{d \nj} = 
  - b\, \frac{ \nj^2 - a_1 \nj + a_0^2}{\nj}  \; .
\label{eq:heavyfit}
\end{equation}
The two relevant fit parameters for the normalized distributions shown
in Figure~\ref{fig:scale_heavy} correspond to the maximum at $\nj =
a_0$, and the (staircase) scaling parameter for QCD jet radiation at
large $\nj$ given by $R = \exp(-b)$. Because we do not include higher
suppression terms towards large $\nj$ we stop the fit at the endpoints
of the curves shown in Figure~\ref{fig:scale_heavy}.

In Table~\ref{tab:scale_heavy} we list the best fit values for these
parameters for both top decays. We immediately see more quantitatively
than in Figure~\ref{fig:scale_heavy} that for example hadronically
decaying top pairs on average include not even one more jet than the
(semi-)leptonic sample. Typically only one of two jets from the $W$
decay is accounted for because of the cutoff at $\ptmin =
50~\gev$. Comparing this value to the $W$ mass it is likely that one
of the two $W$ decay jets gets boosted above $\ptmin$, but the other
one stays below. In contrast, the Jacobian peak of the $b$-quark
energy from the top decay lies above $\ptmin$. Going back to
Table~\ref{tab:scale_heavy} this means that for top pairs the most
likely number of radiated jets is zero, closely followed by one jet
emission~\cite{sgluons}.\bigskip

\begin{table}[t!]
\begin{tabular}{c|cc||ccc|c||ccc}
\hline
& $(t\bar t)_{h h}$ & $(t\bar t)_{\ell \ell/h}$ 
& $\tilde{q}\tilde{q}$ & $\tilde{q} \tilde{g}$   & $\tilde{g} \tilde{g}$ & SUSY 
& $\tilde{q}\tilde{q}$ & $\tilde{q} \tilde{g}$   & $\tilde{g} \tilde{g}$ \\
\hline
& \multicolumn{2}{c||}{full} & \multicolumn{4}{c||}{full} & \multicolumn{3}{c}{jet radiation} \\
\hline
  $a_0$ &  3.13   & 2.34 &  2.89 & 3.53 &  4.16 &  3.15 & n.a. & n.a. & n.a  \\
  $a_1$ &  5.41   & 3.73 &  5.28 & 6.16 &  7.15 &  5.48 &  0.45 & 0.36 & 0.21 \\
  $b$     &  1.25   & 1.07 &  1.71 & 1.25 &  1.09 &  1.27 &  1.14 & 1.07  & 0.98 \\ 
\hline
\end{tabular}
\caption{Parameters defined in Eq.(\ref{eq:heavyfit}) and extracted
  from the unsubtracted distributions shown in
  Figure~\ref{fig:scale_heavy}. The parameter $a_0$ corresponds to the
  position of the maximum while $b$ captures the approximate scaling
  at larger $\nj$. The combined supersymmetric result is based on the
  appropriately weighted event samples for squarks and gluinos.}
\label{tab:scale_heavy}
\end{table}

For squarks and gluinos the features we see in top-pair production
become more pronounced and the jet multiplicity reflects the color
charge of the produced particles. As a reference point in
supersymmetric parameter space we consider reasonably low mass gluinos
and squarks in the SPS1a benchmark scenario~\cite{sps}, with
$m_{\tilde g} = 608~\gev$ and typical light-flavor squarks around
$m_{\tilde q} \sim 558~\gev$.  The new LHC exclusion limits are right
at the edge of excluding this standard parameter choice\footnote{Due
  to the presentation of the LHC results in the $m_0$ vs $m_{1/2}$
  plane it is also not possible to precisely read off the actual limits
  in terms of physics mass parameters. Moreover, since squark and
  gluino masses are both mostly driven by $m_{1/2}$, there does not
  exist a mapping of the $m_0$-$m_{1/2}$ plane into the squark-gluino
  mass plane. Models with significantly heavier gluinos than quarks
  are excluded in CMSSM searches.}.  Because the gluino cannot decay
to a gluon it requires two quarks to get rid of its color
charge. Squark pairs, including squark-antisquark production, predict
two hard decay jets plus some QCD radiation and sub-leading decay
jets. In Table~\ref{tab:scale_heavy} we see that for this production
channel the maximum of a continuous $\nj$ distribution indeed resides
almost at $\nj = 3$. For associated squark-gluino and gluino-pair
production the number of jets increases by almost one, corresponding
to the second gluino-decay jet which not in all cases is hard enough
to appear after requiring $\ptmin =50~\gev$. The jet multiplicity of
the entire supersymmetric sample is close to the average for squark
pair production and squark-gluino production which reflects the
hierarchy in cross sections of the three processes~\cite{prospino}.

Breaking down the supersymmetric signal into individual production
processes we can examine the distinct radiation patterns. Gluino pairs
radiate significantly more than associated production or squark pairs,
which is reflected in the right columns of
Table~\ref{tab:scale_heavy}:
$b(\tilde{g}\tilde{g})<b(\tilde{q}\tilde{g})<b(\tilde{q}\tilde{q})$.
The scaling parameter $R = \exp (-b)$ is consistently larger than for
the background samples in Table~\ref{tab:ratios}. For example for the
jet radiation off squark pair production we find $R \approx
0.32$. Moreover, in Figure~\ref{fig:scale_heavy} we see that the jet
rates for QCD radiation drop off even faster for large
multiplicities. This means that there definitely does not exist any
staircase scaling behavior for heavy particle pair production above a
threshold of 1~TeV at the LHC with a hadronic center-of-mass energy of
7~TeV. This phase space argument should not be mixed with the fact
that the hard scale of such processes and with it the logarithmic
enhancement for collinear radiation is large, \ie the validity of the
collinear approximation extends to larger values of $p_{T,j}$.\bigskip

\begin{figure}[t!]
\includegraphics[width=0.49\textwidth]{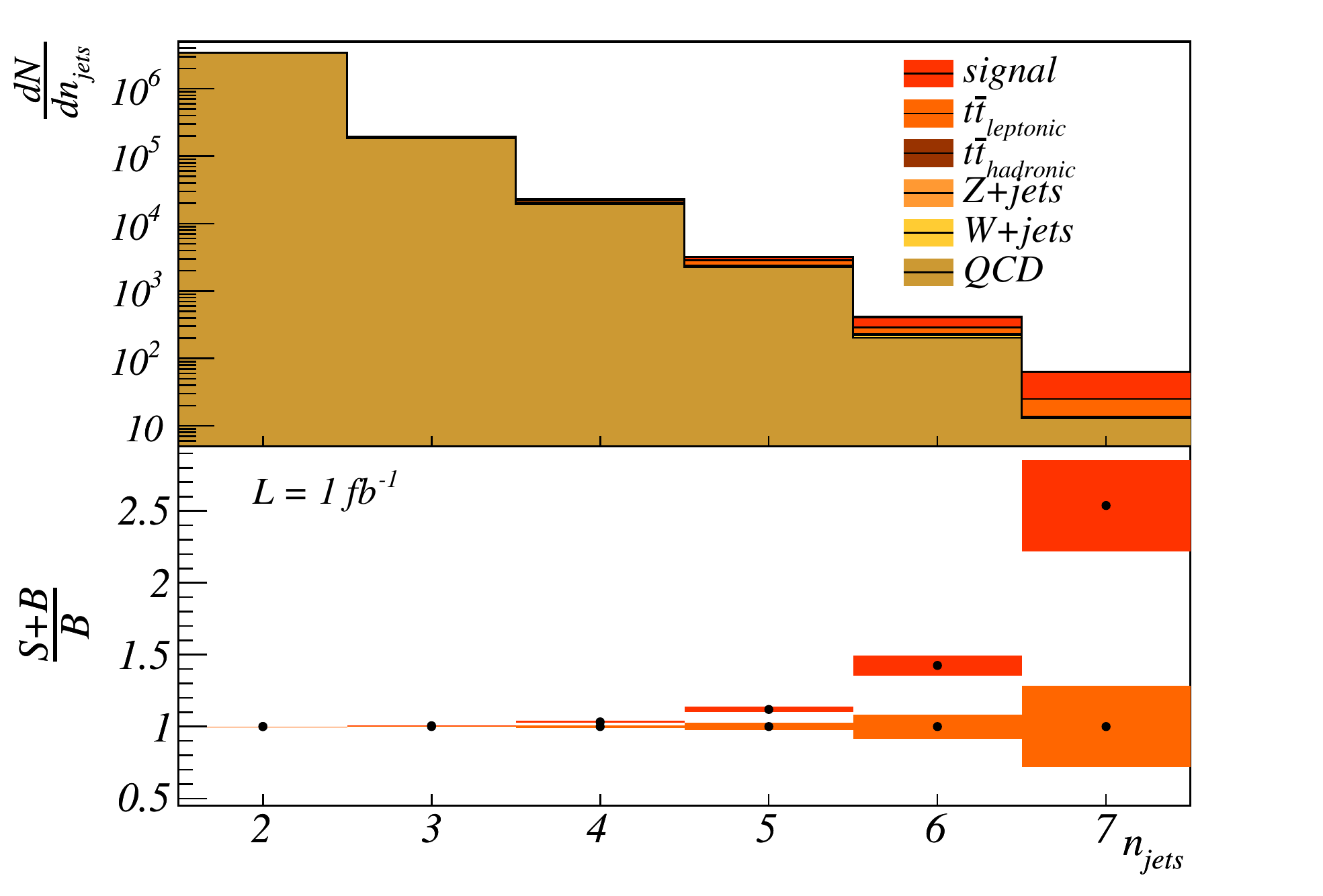}
\caption{\label{fig:njet_sb} Exclusive $\nj$ distribution for all
  considered Standard-Model backgrounds and the SPS1a signal for
  supersymmetry. We present the results for an LHC center-of-mass
  energy of $7$~TeV with an integrated luminosity of $1~\ifb$ and
  after the cuts specified in Eqs.~(\ref{eq:met}) and
  (\ref{eq:jetcuts}).}
\end{figure}

Finally, in Figure~\ref{fig:njet_sb} we show the $\nj$ distribution for
the supersymmetric signal assuming the SPS1a parameter point and the
various Standard Model backgrounds. We apply the background rejection
cuts specified in Eqs.(\ref{eq:met}) and (\ref{eq:jetcuts}). The
variation in shape when including the signal events is statistically
significant and appears as an excess of high jet-multiplicity events
for $\nj > 5$.  The associated statistical significance we compute in
Section~\ref{sec:autofocus}.

\section{Effective mass}
\label{sec:meff}

Before we turn to exploit the number of jets to extract a new physics
signal at the LHC an obvious question is if we can make use of our
understanding of the $\nj$ distribution looking at other observables
in multi-jet final states. More specifically, we will use the measured
scale parameter $\mu/\mu_0$ shown in Figure~\ref{fig:scale_light} to
reliably predict observables, which, based on traditional QCD
simulations, show an overwhelming theory uncertainty.  A classic
observable in this respect is the effective mass~\cite{alan_chris},
which for exclusive jet multiplicities we define as
\begin{equation}
\meff = \met + \sum_\text{all jets} p_{T,j} \;, 
\label{eq:meff}
\end{equation}
including all jets fulfilling Eq.(\ref{eq:jetcuts}).  This definition
is neither optimized to take into account a correlation between hard
jets and the missing-energy vector nor to remove hard initial-state
radiation. Instead, Eq.(\ref{eq:meff}) makes a minimal set of
assumptions to avoid sculpting the background distribution.\bigskip

\begin{figure}[t!]
\includegraphics[width=0.49\textwidth]{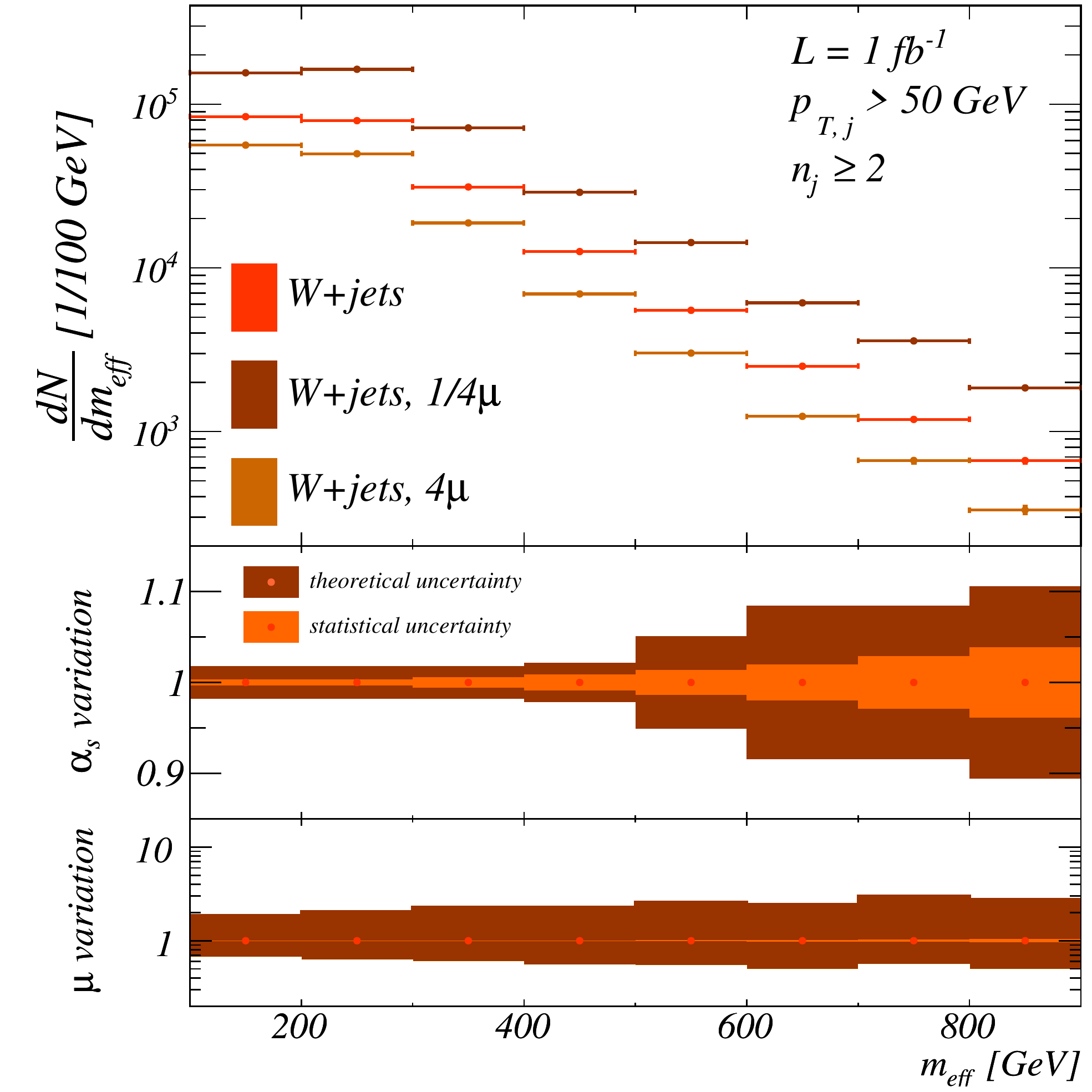}\hfill
\includegraphics[width=0.49\textwidth]{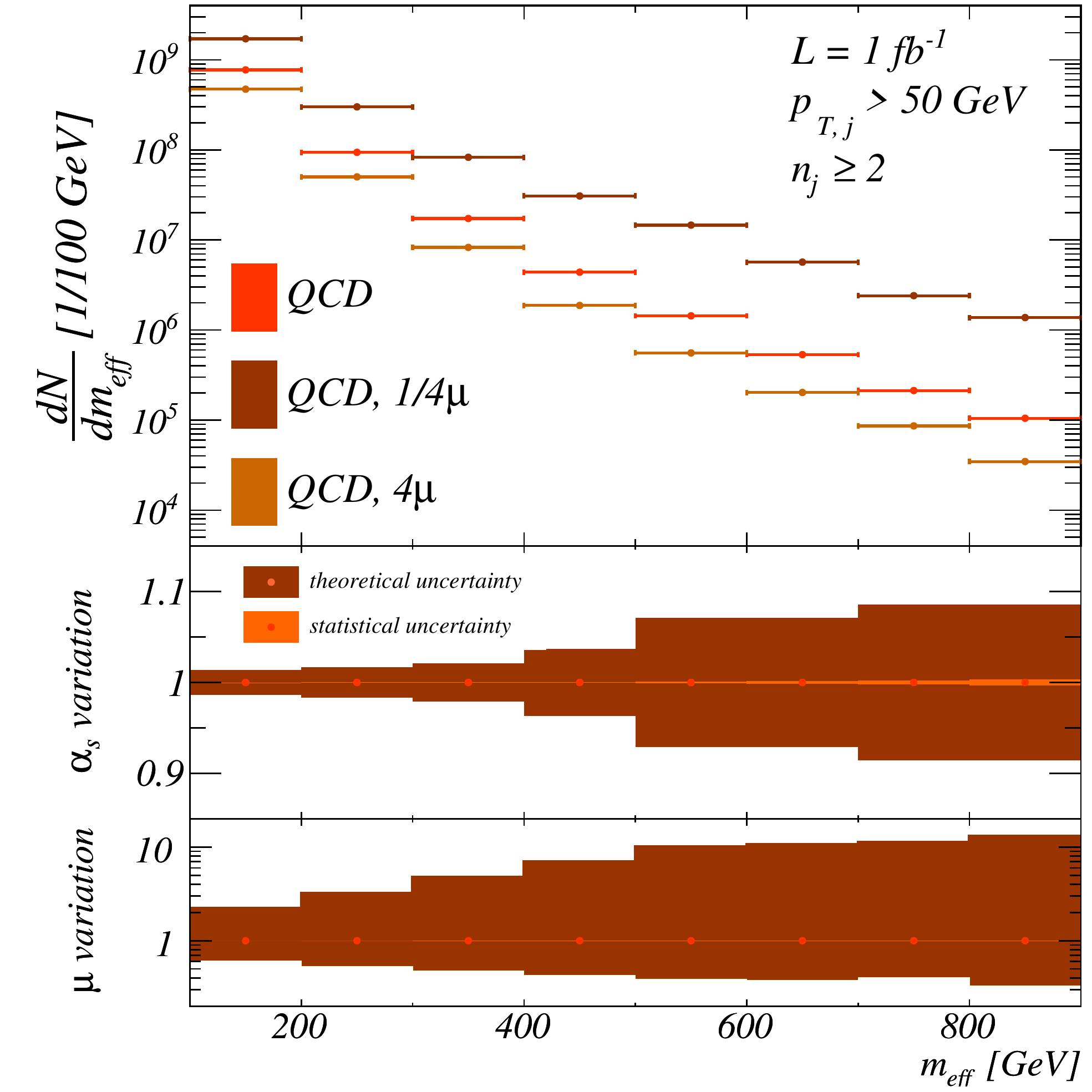}
\caption{Effective mass distribution for $W$+jets and QCD jets
  production. Only the jet cuts given in Eq.(\ref{eq:jetcuts}) are
  applied. The second panels show the parametric uncertainty due to a
  consistent change of $\alpha_s(m_Z)$ between 0.114 and 0.122. The
  third panels show a consistent scale factor variation which can be
  experimentally constrained and should not be considered a theory
  uncertainty.}
\label{fig:meff} 
\end{figure}

Just like the $\nj$ distribution $\meff$ in the Standard Model cannot
be reliably predicted by parton-shower Monte Carlos.  Jets entering
the sum in Eq.(\ref{eq:meff}) we have to understand over their entire
transverse-momentum spectrum.  {\sc Ckkw}~\cite{ckkw,mets_matching} or
{\sc Mlm}~\cite{mlm} matching is therefore the most adequate approach
for simulating $\meff$.

Exactly following the treatment of the $\nj$ distribution in
Section~\ref{sec:jetcount} we estimate two sources of theory
uncertainties, the parametric error from varying the strong coupling
and the scale-variation systematics. To not be limited by statistics
of our background samples we for now discard the missing energy cut
and the lepton veto and instead study the fully inclusive
processes. In Figure~\ref{fig:meff} we present the $\meff$
distribution for $W+$jets and the QCD jets production with $\nj\geq
2$. The same way as in Figure~\ref{fig:scale_light} we show the
relative impact of the two sources of uncertainty in the lower
panels. The parametric error from $\alpha_s(m_Z)$ ranges well below
20\% even towards large values of $\meff$. For the
electroweak process this is of similar size to the expected
statistical error for an integrated luminosity of $1~\ifb$. As
expected, towards large $\meff$ the error band increases, but not
dramatically.

In contrast, the scale-factor variation $\mu/\mu_0 = 1/4 - 4$ has
a huge effect on the $\meff$ simulation, essentially rendering it
unpredictive. For values above $\meff = 500~\gev$ the error bands
become large enough to make it impossible to extract new physics from
this observable, were we to consider the scale variation a proper
theory error. However, measurements of multi-jet rates and other jet 
observables at Tevatron and LHC indicate that for the case of {\sc Sherpa}
this scaling factor is approximately one~\cite{sh_validation}. Measuring 
the staircase-scaling factors even more precisely with the 2011 LHC data 
will further constrain the scale ambiguities underlying our 
QCD simulations -- allowing us to make reliable predictions for \eg the 
$\meff$ observable.\bigskip

\begin{figure}[b!]
\includegraphics[width=0.49\textwidth]{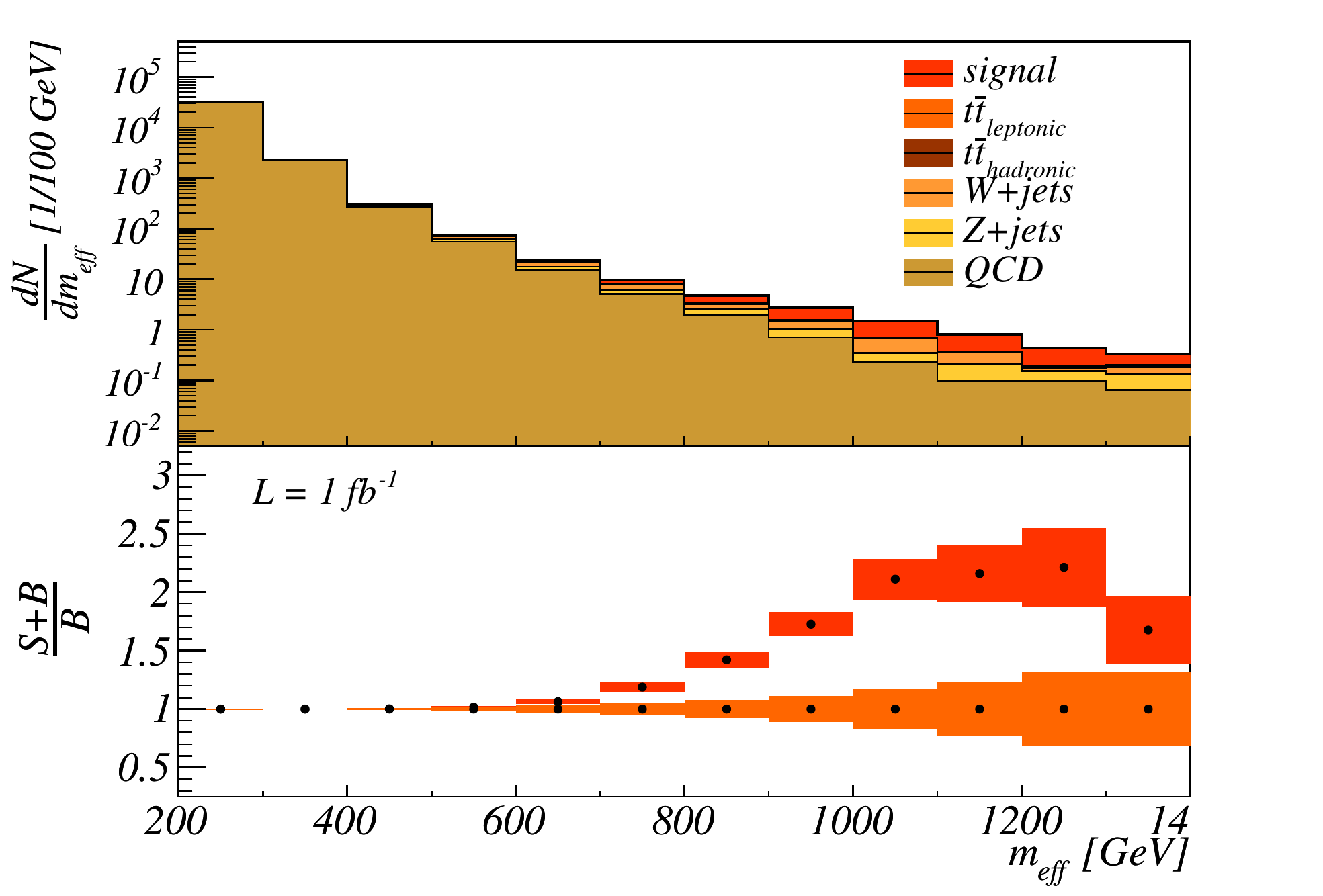}\hfill
\includegraphics[width=0.49\textwidth]{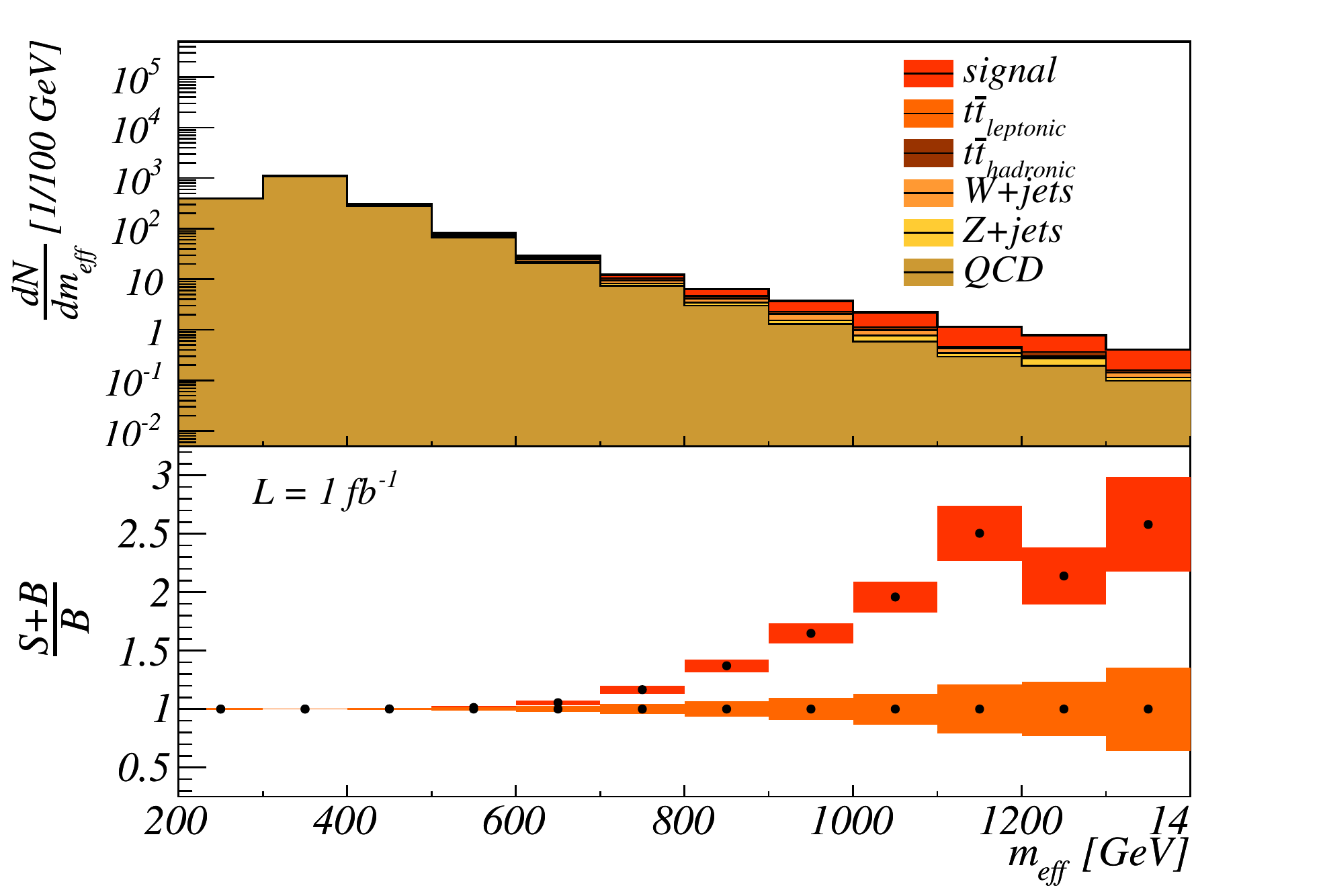}
\caption{\label{fig:meff_sb} Effective mass distribution for exclusive
  2-jet and 3-jet events for Standard-Model backgrounds and the
  supersymmetric signal using the SPS1a parameter point. We assume a
  center-of-mass energy of $7$~TeV with an integrated luminosity of
  $1~\ifb$ and apply all cuts in Eqs.(\ref{eq:met}) and
  (\ref{eq:jetcuts}).}
\end{figure}

To see the impact of $\meff$ in searches for supersymmetry we show the
$\meff$ distribution for exclusive $2$-jet and $3$-jet events in
Figure~\ref{fig:meff_sb}. It includes Standard-Model backgrounds as well
as the supersymmetric signal. All jet-selection and background-rejection 
cuts specified in Eqs.(\ref{eq:met}) and (\ref{eq:jetcuts})
are applied. As mentioned before, QCD jets are the by far dominant
channel. Only for $\meff > 800$~GeV the signal starts overcoming the
backgrounds.  The statistical uncertainty for $1~\ifb$ we indicate by
the shaded regions in the lower panels. It is worth noticing that the
signal+background sample when compared to the pure background sample
exhibits a maximum at around $\meff \sim 1.1$~TeV.  This scale
corresponds to the squark and gluino masses which for pair production
add to 1100 to 1200~GeV.  This means that the $\meff$ distribution for
exclusive jet multiplicities can serve as background rejection as
well as a measure for the mass scale of the new heavy colored states.

\section{Autofocusing}
\label{sec:autofocus}

Following our results in the previous sections we should be able to
use the shapes of the $\nj$ and $\meff$ distributions to extract a
supersymmetric signal from the now quantitatively understood Standard
Model backgrounds. Given that the two distributions are affected
independently by the color structure of the new physics sector and by
its mass scale(s) we will assess the power of the two-dimensional
$\nj$ vs $\meff$ correlations in extracting a discovery or an
exclusion. Such a two-dimensional shape analysis is the natural second
step after the first completely inclusive searches based on counting
events.  According to Sections~\ref{sec:jetcount}-\ref{sec:meff}
systematic experimental uncertainties will start dominating for
luminosities around ${\cal{O}}(1~\ifb)$.  Since those are subject to
continuous refinement during data taking and need to be addressed
within a full detector simulation study we limit ourselves to statistical
uncertainties for a given luminosity. While this means that we will
not obtain reliable estimates for the discovery reach, we will see that 
it allows us to discuss the main benefits and limits of the proposed
analysis.\bigskip

As supersymmetric reference models we choose the benchmark point
SPS1a, two variations of it, and SPS4.  Again, we only apply the cuts
given in Eqs.(\ref{eq:met}) and (\ref{eq:jetcuts}) and use the
exclusive definition of $\nj$ and $\meff$.  For the $\meff$
distribution we choose a binning of 100~GeV, which approximately
reflects the experimental resolution towards large $\meff$.

For given background and signal+background hypotheses we use a binned
log-likelihood ratio to compute statistical significances assuming
statistically uncorrelated bins
\begin{equation}
 \log Q 
=  \, \sum_\text{bins}
   \left [ n_i\log\left(1+\frac{s_i}{b_i} \right)- s_i 
   \right ] \; .
\label{eq:loglike1}
\end{equation}
It includes the luminosity via the signal and background event numbers
$s_i$ and $b_i$ in each bin. While it avoids the limitations of
$S/\sqrt{B}$ in regions requiring Poisson statistics it approaches a
Gaussian limit for each individual channel when the bin content
becomes large.  Some features of this well established approach we
summarize in Appendix~\ref{app:hypo}.  Applying a ``simple hypothesis
test'' tells us how likely it is that the background-only hypothesis
fakes the predicted signal+background distributions as a statistical
fluctuation, \ie we define the $p$-value as the SPS1a likelihood
ratio's median.  The likelihood ratio given in Eq.(\ref{eq:loglike1})
we compute for the exclusive $\nj$, $\meff$, and two-dimensional
$(\nj,\meff)$ distributions. In this two-dimensional plane the
definition of $\meff$, following Eq.(\ref{eq:meff}), only includes
exactly $\nj$ jets. With this completely exclusive definition of $\nj$ 
and $\meff$ we ensure that the sum over all bins in the $(\nj,\meff)$
reproduces the total cross section.\bigskip

Considering this correlation is similar in spirit to the $(\met,H_T)$
analysis proposed in Ref.~\cite{jay_corr}. However, first we focus on
the $\nj$ and $\meff$ distributions because in
Sections~\ref{sec:jetcount}-\ref{sec:meff} we have shown that we can
quantitatively understand the staircase scaling behavior of the
Standard Model backgrounds and translate its precision into other
variables. In addition, as we will see in this section these two
variables play a special role, as they not only distinguish signals 
from backgrounds, but also contain information on the structure of the
underlying new-physics model. As mentioned above, for the sake of a
proof of concept we ignore all uncertainties except for statistical
experimental errors, to avoid correlations in the definition of the
log-likelihood.\bigskip

\begin{table}[t]
\begin{tabular}{l|c}
\hline
& {signal significance} \\ & for $35~\ipb$\\
\hline
inclusive &   $0.2\,\sigma$\\
\hline 
$n_{\text{jets}}$ (1D) &   $1.6\,\sigma$ \\
$m_{\text{eff}}$ (1D)  &    $3.3\,\sigma$ \\
\hline
$( n_{\text{jets}}, m_{\text{eff}})$ (2D) &     $ 4.6\,\sigma $  \\
\hline
\end{tabular}
\caption{\label{tab:binning}Confidence levels for the signal plus
  background sample ruling out the background-only hypothesis based on
  one and two dimensional log-likelihood distributions. The
  supersymmetric mass spectrum is given by SPS1a.}
\end{table}

We can expect from Figures~\ref{fig:njet_sb} and \ref{fig:meff_sb}
that the rate in each individual $\nj$ bin is dominated by Standard-Model
processes at low $\meff$. Most likely, this region will be the control
region to normalize the QCD and $W/Z$+jets backgrounds. With the
exception of hadronically decaying top pairs all Standard-Model
channels will then show a simple decrease in both directions of the
two-dimensional $(\nj,\meff)$ plane which we can predict following the
arguments in Sections~\ref{sec:jetcount}-\ref{sec:meff}.  The signal
contribution will become visible only once $\meff$ reaches the mass
range of the particles produced.

In Table~\ref{tab:binning} we compare the statistical significances
for the supersymmetric SPS1a parameter point at 7 TeV center-of-mass
energy for the various analysis strategies: first, we show the results
based on the total production rates after the inclusive cuts of
Eqs.(\ref{eq:met}) and (\ref{eq:jetcuts}). As expected, including the
signal events leaves us completely consistent with the background-only
hypothesis. Next, the likelihood ratio computed from the $\nj$
distribution gives rise to sizable deviations from the background for
integrated luminosities as small as $35~\ipb$. The one-dimensional
$\meff$ distribution turns out to be an even better discriminator. It
gives us more than twice the $\nj$ significance, namely $3.3\,\sigma$
for ${\cal{L}}=35~\ipb$. The highest significant discriminative power
we obtain for the two-dimensional binned $(\nj,\meff)$ case. This is a
direct consequence of the additive binned log-likelihood given in
Eq.(\ref{eq:loglike1}).\bigskip

\begin{figure}[b!]
\begin{center}
  \includegraphics[width=0.24\textwidth]{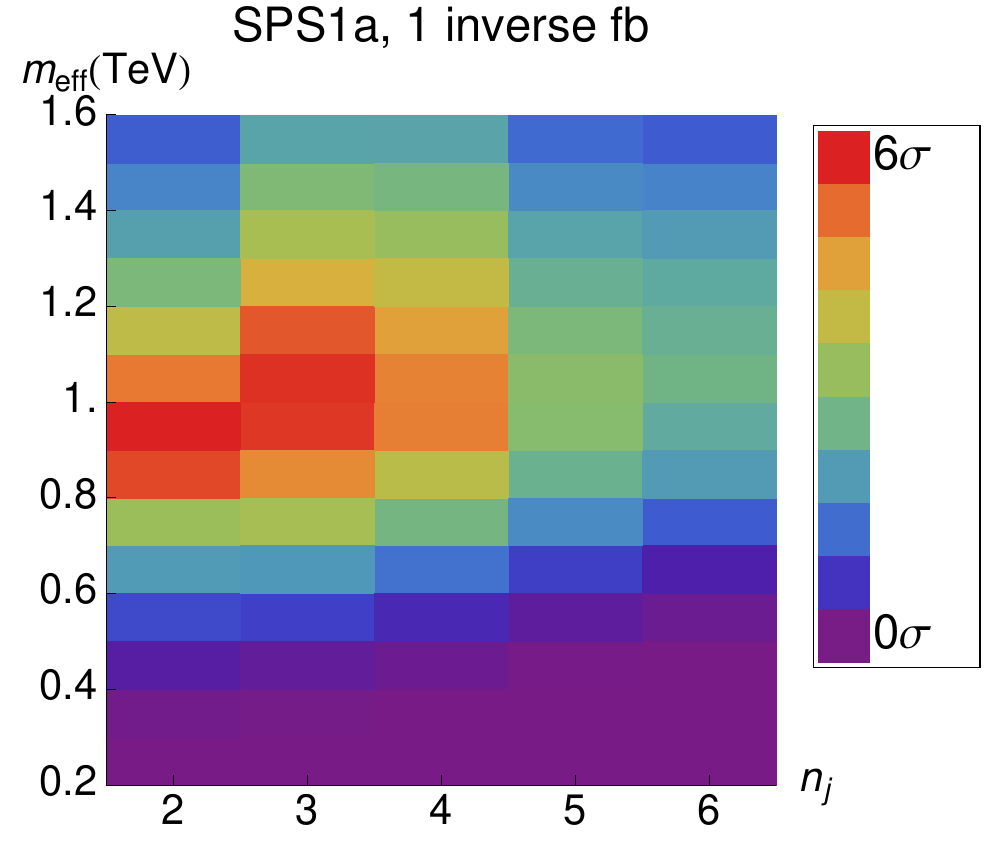}
  \includegraphics[width=0.24\textwidth]{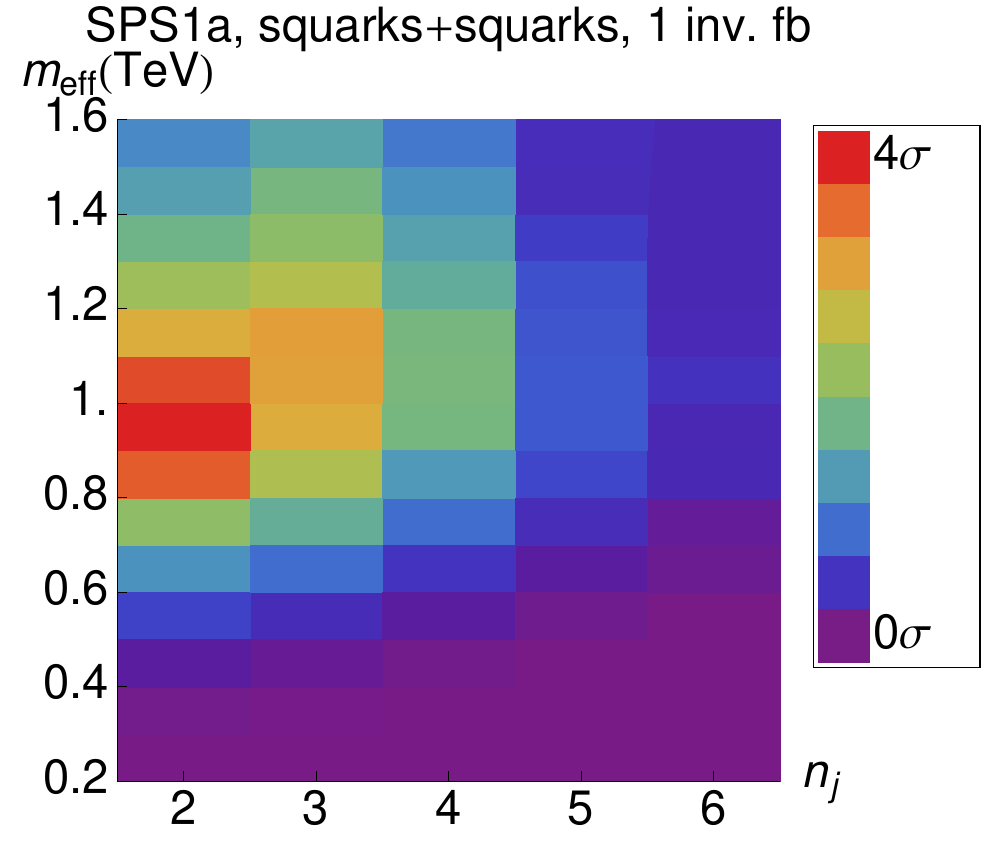}
  \includegraphics[width=0.24\textwidth]{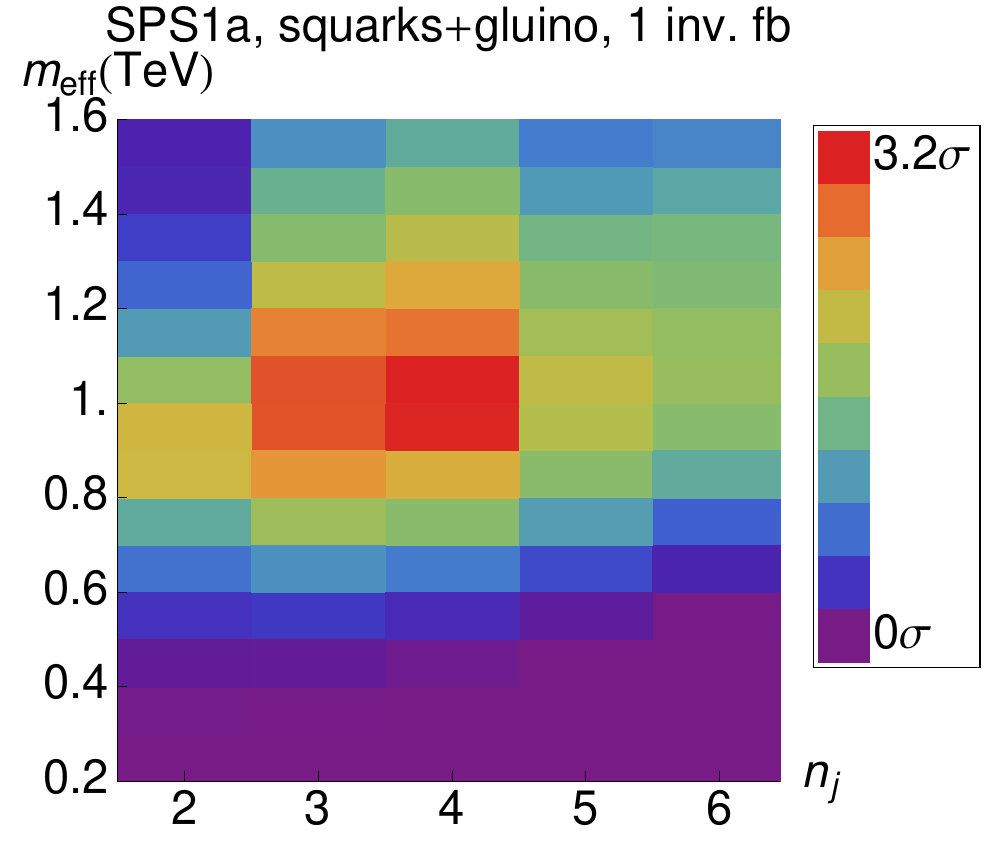}
  \includegraphics[width=0.24\textwidth]{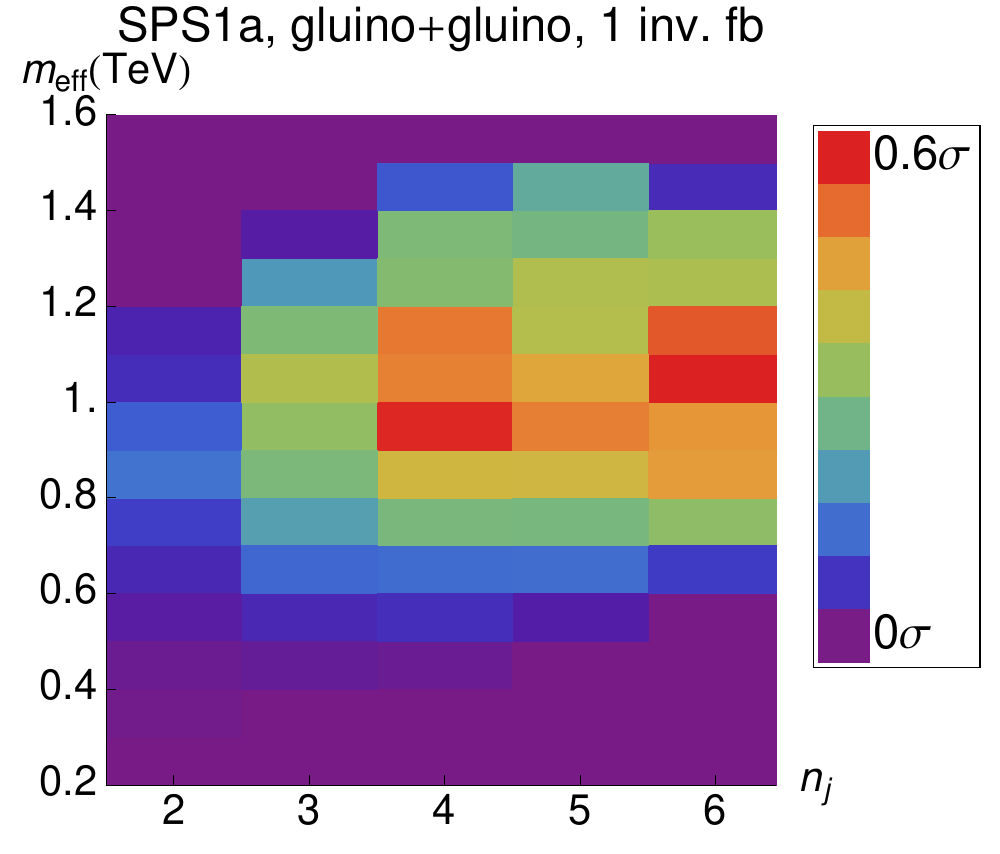}
\end{center}
  \caption{\label{fig:llhrplanesps1} Log-likelihood contributions over
    the $(\nj,\meff)$ plane for the supersymmetric signal using the
    SPS1a spectrum.  The color code is normalized to different maximum
    significances.}
\end{figure}

Beyond the relevance of the $(\nj,\meff)$ distributions to extract new
particles from backgrounds, we can utilize it to study signal
properties.  Above, we argue that new physics contributions to $\nj$
will only appear once $\meff$ reaches the mass scale of the sum of
both heavy particles produced. However, this only happens if the
exclusive $\nj$ value allows us to include the decay jets contributing
to $\meff$. Hence, the new physics contributions to the two
observables will show a correlation based on the mass and decay
channels of the new particles produced. The decay channels can
typically be linked to the color charge of the new particles if we
assume that the missing energy particle cannot carry color
charge. Color triplets will tend to decay to one hard quark jet while
color octets with their diagonal coupling to gluons will radiate two
quark jets.  This means breaking down the binned log-likelihood ratio
over the fully exclusive ($\nj,\meff$) plane and keeping track of the
individual contribution of each bin will automatically focus our
search on the appropriate properties of the particles we are looking
for.\bigskip

This statement is not limited to supersymmetry, the SPS1a parameter
point or any other assumption about the signal. It can be applied to
general physics beyond the Standard Model with strongly interacting
new particles and a stable dark matter candidate.  In
Figure~\ref{fig:llhrplanesps1} we show the contributions of the
individual bins to the summed log-likelihood ratio for all signal
events combined and split into three production processes.  The
maximum significance automatically reflects SPS1a's decay paradigm
$\tilde q \tilde q^{(*)} \rightarrow 2$~jets and $\tilde q^{(*)}
\tilde{g} \rightarrow 3$~jets, and $\tilde{g}\tilde{g}\rightarrow
4$~jets, know already from Figure~\ref{fig:scale_heavy}. The first two
channels we can study using an integrated luminosity of
$1~\ifb$. Squark pair production is dominant because at the LHC it
includes a quark-quark initial state. Associated production, which
often is the dominant channel at the LHC, has a comparable statistical
yield and features a slightly higher $\meff$ range. Both channels
combined define the diagonal correlation we see for the combined
signal events.

Gluino pair production has the smallest production rate and therefore
becomes subleading in the combined supersymmetry sample. However, for
this channel we can best follow the imprint of higher jet
multiplicities.  Due to its large mass and its color charge gluino
pairs produce significantly more jet radiation which we can resolve
for a sufficiently low $\ptmin$ threshold. For $\ptmin = 50$~GeV we
might just capture the first decay jet from the gluino cascade,
reflecting the mass hierarchy $m_{\tilde{g}} - m_{\tilde{q}} \sim
60$~GeV.  The peak in the log-likelihood plane around $\nj=4$ results
from the maximum in the $\tilde{g}\tilde{g}$ production cross
section. For $\nj=5$ the background is still large compared to the
signal, but dropping at an exponential rate it gets surpassed for
$\nj=6$, explaining the structure we observe in
Figure~\ref{fig:llhrplanesps1}.\bigskip

\begin{figure}[!t]
\begin{center}  
  \includegraphics[width=0.24\textwidth]{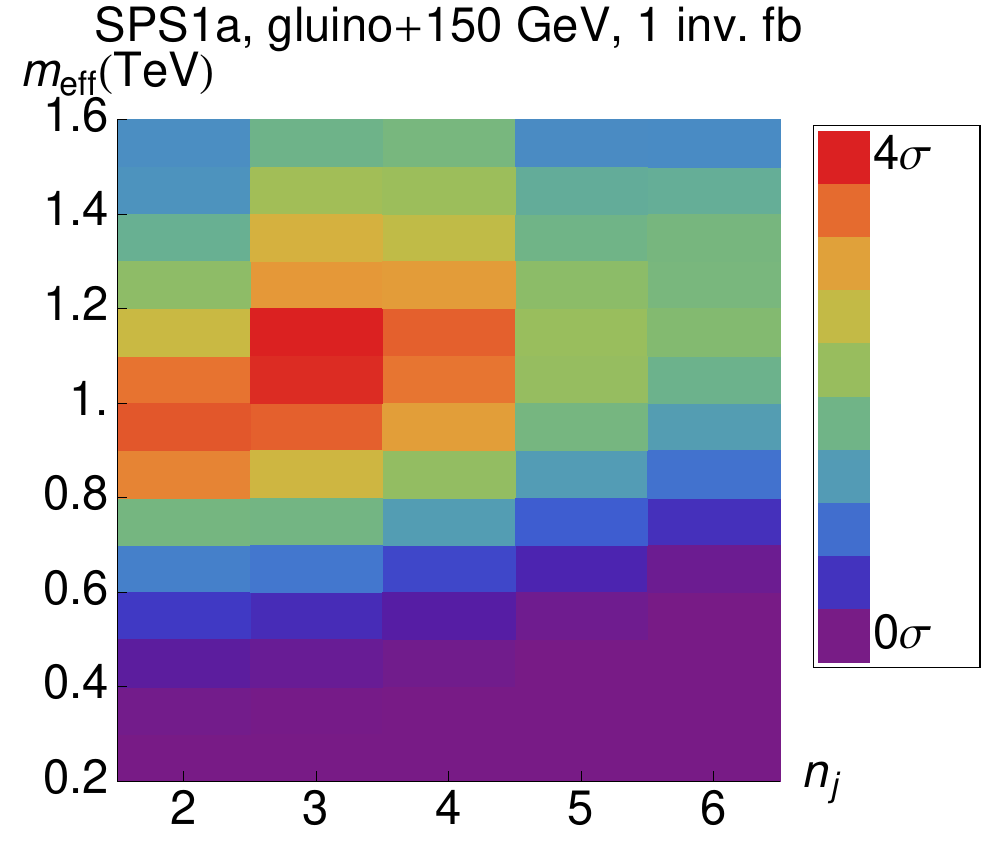}\hfill
  \includegraphics[width=0.24\textwidth]{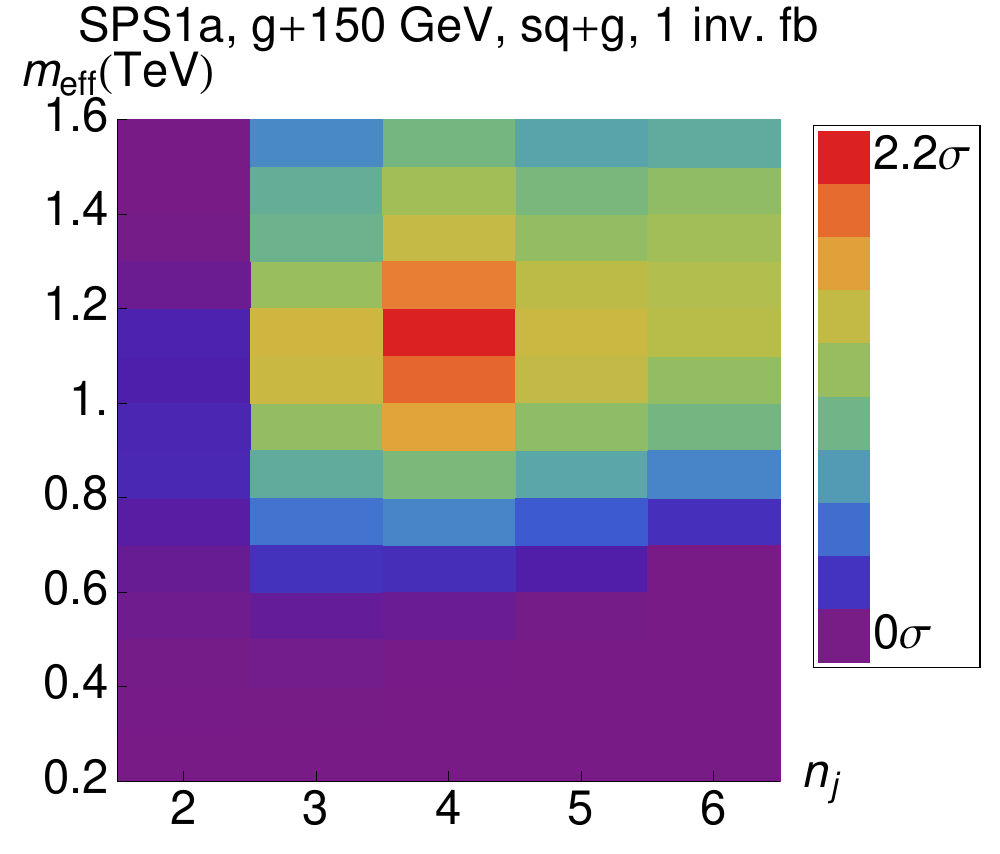}\hfill
  \includegraphics[width=0.24\textwidth]{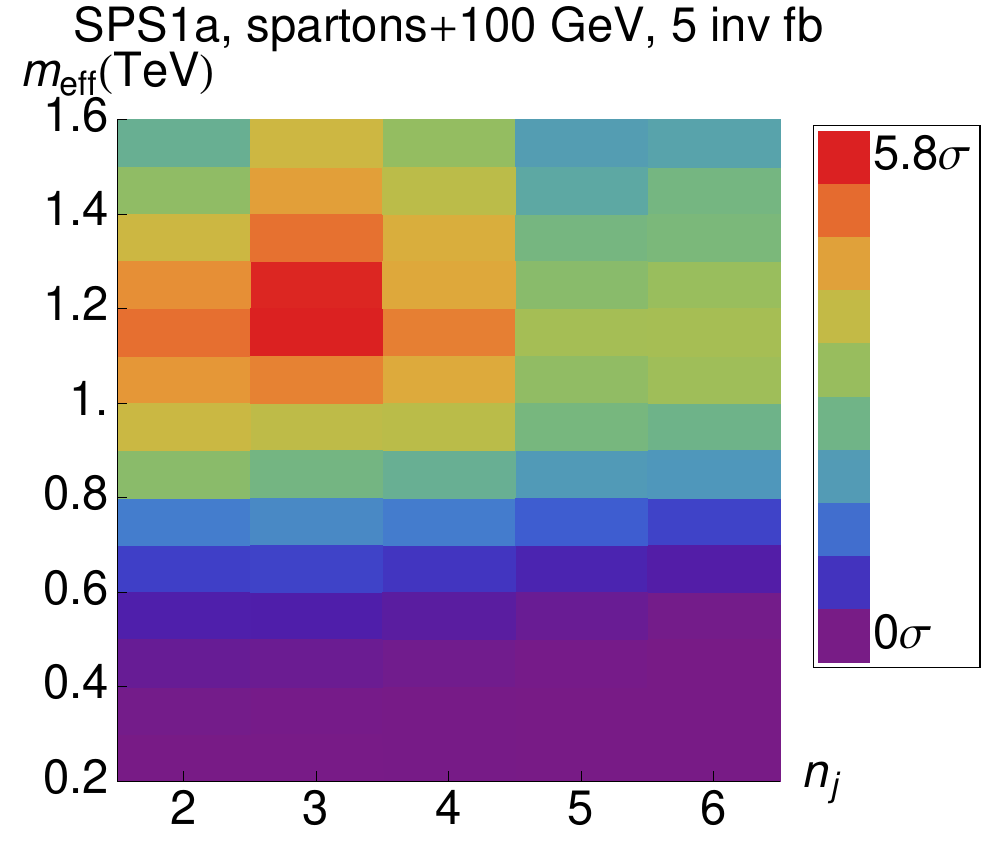}\hfill
  \includegraphics[width=0.24\textwidth]{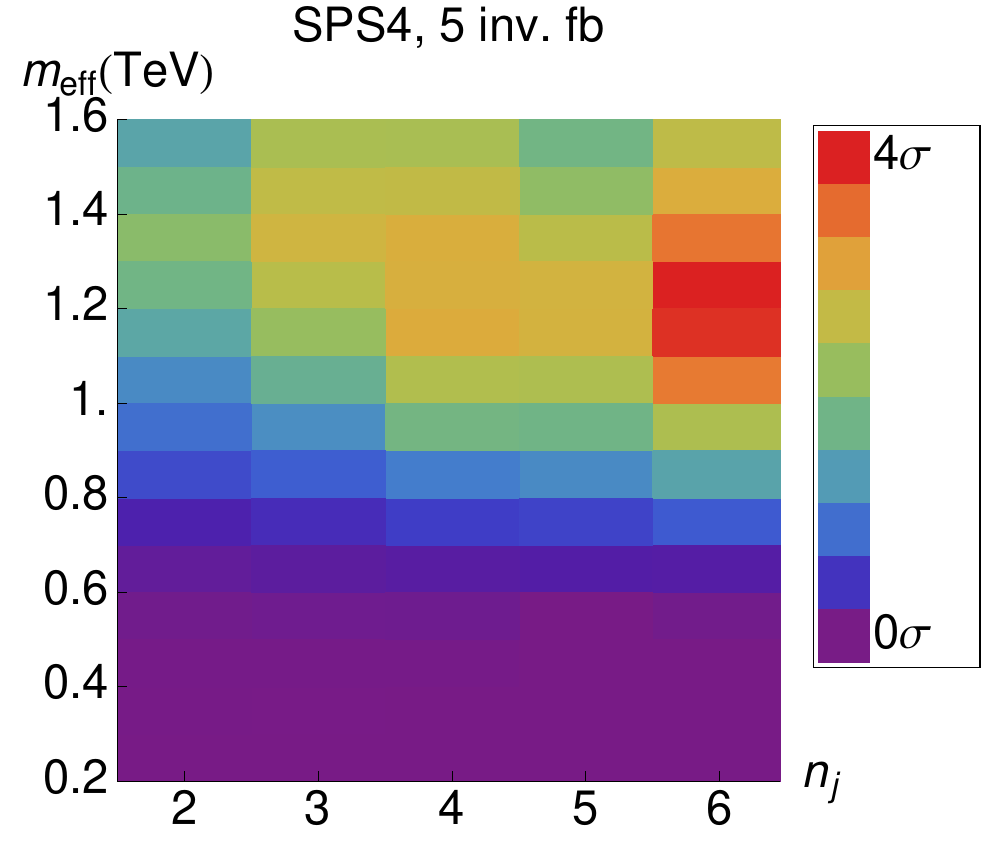}
\end{center}
  \caption{\label{fig:llhrplane} Log-likelihood distributions over the
    $(\nj,\meff)$ plane for the different supersymmetric spectra. Due
    to the smaller signal rates we present results for an integrated
    luminosity of $5~\ifb$.}
\end{figure}

Finally, we can study how changes to the new physics spectrum are
reflected in the significances computed from the binned log-likelihood
$Q(\nj,\meff)$. We investigate three different supersymmetric mass
spectra : first, we increase only the gluino mass by 150~GeV with
respect to SPS1a ($\sigma^{\text{NLO}}_\text{SUSY}=2.69~\pb$ according to {\sc
  Prospino2.1}~\cite{prospino}); second, we increase all
colored-sparticle masses by 100~GeV with respect to SPS1a
($\sigma^{\text{NLO}}_\text{SUSY}=1.63~\pb$); third, we consider the SPS4
benchmark~\cite{sps} with an inverted mass hierarchy $m_{\tilde q}\sim
750~\gev > m_{\tilde{g}} \sim 730~\gev$
($\sigma^{\text{NLO}}_\text{SUSY}=0.83~\pb$).  All of these cross sections are
significantly smaller than for SPS1a with its
$\sigma^{\text{NLO}}_\text{SUSY}=4.68~\pb$, which means we increase our nominal
luminosity to ${\cal{L}}=5~\ifb$.

In Figure~\ref{fig:llhrplane} we clearly see the effect of the
increased gluino mass. The $\meff$ peak for associated squark-gluino
production moves to larger values, as does the $\nj$ maximum. However,
because the balance between squark pair production and associated
squark-gluino production shifts into the direction of the squark pairs,
this effect is not quite as pronounced.  The second scenario with
increased squark and gluino masses leads to a pronounced maximum at
larger $\meff$.  Due to the smaller signal cross section the
sensitivity in particular in the $\nj=2$ bins gets considerably
diminished, appearing as a shift towards higher $\nj$ values.  For the
all-hadronic search in the SPS4 parameter point longer decay chains
for gluinos through bottom squarks appear in the high $\nj$ bins
only.\bigskip

The SPS4 case illustrates that $\nj =6$ does not have to be the
maximum jet multiplicity we need to consider. Once we rely on a
combination of data and Monte Carlo methods to describe the $\nj$
staircase scaling for background processes we can extend our analyses
to very large jet multiplicities. On the other hand,
Figure~\ref{fig:llhrplanesps1} also clearly indicates that for example
in the SPS1a parameter point the optimal signal extraction strategy by
no means requires us to go to very large jet multiplicities. For the
SPS1a parameter point the two-jet bins are leading contributions to
the total significance.

\section{Outlook}

Multi-jet events with and without a weak gauge boson are the dominant
backgrounds to inclusive searches for example for
supersymmetry. Simulating them with traditional parton shower Monte
Carlos does not lead to reliable results. This changes when we make
use of jet merging to predict the shapes of multi-jet backgrounds.\bigskip

One of the most challenging distributions to describe theoretically is
the inclusive or exclusive number of jets per event. From data we know
that the inclusive $W/Z$+jets distributions without rigid cuts follow
a {\sl staircase scaling} with constant jet ratios $R =
\hat{\sigma}_{n+1}/\hat{\sigma}_n$. We have shown that staircase
scaling in the inclusive jet rates is equivalent to the same scaling
in exclusive rates. Nowadays, the exclusive scaling we can compare to
the predictions of jet merging Monte Carlos, like {\sc
  Sherpa}. Moreover, we have seen that QCD jet production shows an
even more pronounced scaling than $W/Z$+jets production to jet
multiplicities of $\nj = 8$.\bigskip

We studied the validity of the staircase scaling behavior, the theory
uncertainties in the $\nj$ distributions, its link to other multi-jet
observables, and the application of jet-exclusive observables to new
physics searches and found:
\begin{enumerate}
\item While we cannot derive the staircase scaling of the $\nj$
  distribution from first principles we can reproduce it using the
  appropriate Monte Carlo tools. This includes the scaling feature
  itself, a careful error analysis, and the scaling violation effects
  towards large values of $\nj$ due to phase space restrictions.
\item The theory uncertainty on the staircase scaling consists of
  tunable parameters like scale factors in the factorization and
  renormalization scales and on parametric errors like the dependence
  on $\alpha_s$. The latter are small. The scale factor hugely
  overestimates the error and should be thought of as a tuning
  parameter for the different jet merging implementations. For {\sc
    Sherpa} it comes out close to unity.
\item The scaling parameter $R = \sigma_{n+1}/\sigma_n$ depends on the
  hard process and on kinematic cuts. Both effects we can reliably
  predict using Monte Carlos, as we have shown for the $W/Z$+jets and
  pure jets cases. The $\gamma$+jets case we postpone to a later study
  with more specific details for the LHC experiments~\cite{future}.
\item These simulations of the staircase scaling in multi-jet
  processes can be easily combined with data driven techniques, giving
  us the over-all normalization and a cross check for the first $\nj$
  bins. Statistically limited regions of phase space will become
  accessible via simulations, including a reliable error estimate.
\item Understanding staircase scaling of multi-jet processes allows us
  to predict other multi-jet variables, like the effective mass
  $\meff$. Again, this includes a proper treatment of theory
  uncertainties. In addition, the completely inclusive definition of
  $\meff$ removes dangerous artifacts due to the usual truncations.
\item Based on for example the $\nj$ vs $\meff$ correlation for a
  fixed $\ptmin$ we can define a likelihood-based analysis avoiding
  model or spectrum specific background rejection cuts. Such shape
  analyses in multi-jet search channels are the natural extension of
  the early inclusive ATLAS and CMS searches.
\end{enumerate}
Of course this simple first approximation to the exclusive zero-lepton
search for jets plus missing energy is not the only application of
such methods. Searches including leptons, $b$ tags, or hard photons
will benefit from the same treatment, as long as they include
non-negligible numbers of jets. The same is true for hadronically
decaying top quarks in the Standard Model.

\bigskip
\begin{center}
{\bf Acknowledgments}
\end{center}
\smallskip

We would like to thank Elmar Bittner and Andreas Nussbaumer for
computing support.  All simulations underlying this study have been
performed on bwGRiD (\url{http://www.bw-grid.de}), member of the
German D-Grid initiative, funded by the Ministry for Education and
Research (Bundesministerium f\"ur Bildung und Forschung) and the
Ministry for Science, Research and Arts Baden-W\"urttemberg
(Ministerium f\"ur Wissenschaft, Forschung und Kunst
Baden-W\"urttemberg).

\appendix

\section{Hypothesis test}
\label{app:hypo}

\begin{figure}[!b]
\includegraphics[width=0.43\textwidth]{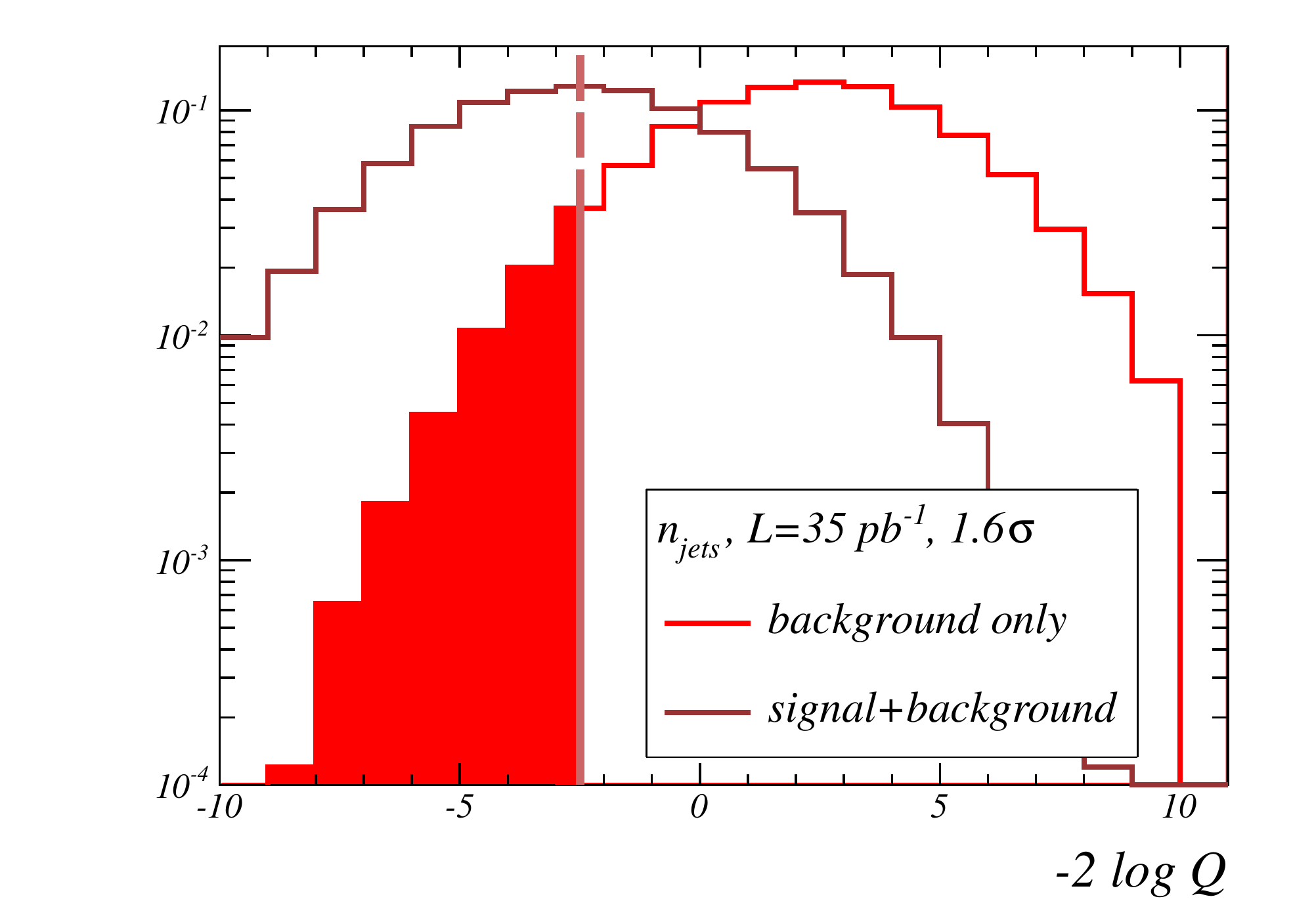}
\caption{Log-likelihood ratio distributions based on the $\nj$
  discriminator for a luminosity of $35~\ipb$.  The confidence level is
  computed by evaluating the overlap of the background-only
  distribution with the signal+background maximum.}
\label{fig:llhrdisc}
\end{figure}

In this section we briefly review the binned log-likelihood ratio
hypothesis tests which we apply in Section~\ref{sec:autofocus}.  It
discriminates between two specific hypotheses and has been used for
the combined LEP-Higgs limits~\cite{lep_higgs}, Tevatron
analyses~\cite{junk}, and in various contexts of LHC Higgs
phenomenology~\cite{nielsen}.  According to
the Neyman--Pearson lemma~\cite{neyman} the likelihood ratio is the
most powerful test statistic (\eg signal+background vs
background-only). We compute the (binned) log-likelihood ratio
\begin{equation}
\label{eq:loglike2}
{\cal{Q}}
= -2 \log Q 
= -2\log \frac{L( \text{data}\, |\, {\cal{S}} + {\cal{B}} ) } 
              { L( \text{data}\,| \, {\cal{B}} ) }
= 2 \, 
    \left[ s - n \log \left(1+\frac{s}{b} \right) 
    \right] 
\stackrel{\text{binned}}{=} 2 \, \sum_{i\in \text{bins}} 
    \left[ s_i - n_i\log \left(1+\frac{s_i}{b_i} \right) 
    \right] \; ,
\end{equation}
where $s$ and $b$ denote the signal $\mathcal{S}$ and background
$\mathcal{B}$ event numbers for a given luminosity, split into the
bins $i$.  The probabilities to observe $n$ events given the expected
numbers $s$ and $b$ in Eq.(\ref{eq:loglike2}) are determined by
Poisson distributions
\begin{equation}
L( \text{data}\, |\, {\cal{S}} + {\cal{B}} ) = \frac{(b+s)^{n} e^{-(s+b)}}{n!}
\qquad \qquad \qquad
L( \text{data}\, |\, {\cal{B}} ) = \frac{b^{n} e^{-b}}{n!} \; .
\end{equation}
The sum in Eq.(\ref{eq:loglike2}) extends over all contributing
channels. The likelihood distributions we generate as pseudo-data
around each hypothesis' central value, which means that in principle
we can include any kind of correlation. In this work we limit
ourselves to statistically independent bins $i$ of the $\nj$ and
$\meff$ distributions. The set of entries in each bin $\{n_i\}$ we
simulate numerically and histogram them as a function of ${\cal{Q}}$,
following the Neyman-Pearson lemma. To simulate the log-likelihood
distributions we need to specify which hypothesis the bin entries
$\{n_i\}$ should follow, \ie we can compute $\qsb$ or $\qbb$.  In
Figure~\ref{fig:llhrdisc} we show both ${\cal{Q}}$ distributions for
the binned one-dimensional $\nj$ distribution studied in our paper.

In our analysis we are interested in the probability that the
background alone fakes the expected signal+background
distributions. This confidence level is given by the integral of the
background distribution $\qbb$ over the signal+background range, indicated
by the red-shaded region in Figure \ref{fig:llhrdisc}. This
signal+background range is defined as all likelihood values above the
median of the likelihood distribution assuming the signal+background
hypothesis
\begin{equation}
\text{CL}_{\cal{B}} 
= \int_{-\infty}^{\qss} \d {\cal{Q}} \, \qbb 
= \mathrm{erfc} \left( \frac{Z}{\sqrt{2}} \right) \; ,
\end{equation}
where for illustration purposes we convert the confidence levels into
the Gaussian number of standard deviations $Z$ via the inverse error
function.


\end{document}

%% file: declare.tex
\def\nj{n_\text{jets}}
\def\meff{m_\text{eff}}
\def\ptmin{p_T^\text{min}}

\newcommand\one{\leavevmode\hbox{\small1\normalsize\kern-.33em1}}

\newcommand{\met}{\slashchar{p}_T}

\newcommand{\gev}{{\ensuremath\rm GeV}}

\newcommand{\pb}{{\ensuremath\rm pb}}

\newcommand{\ifb}{{\ensuremath\rm fb^{-1}}}
\newcommand{\ipb}{{\ensuremath\rm pb^{-1}}}

\def\slashchar#1{\setbox0=\hbox{$#1$}           
   \dimen0=\wd0                                 
   \setbox1=\hbox{/} \dimen1=\wd1               
   \ifdim\dimen0>\dimen1                        
      \rlap{\hbox to \dimen0{\hfil/\hfil}}      
      #1                                        
   \else                                        
      \rlap{\hbox to \dimen1{\hfil$#1$\hfil}}   
      /                                         
   \fi}

\def\eg{{\sl e.g.} \,}
\def\ie{{\sl i.e.} \,}

\renewcommand{\d}{{\mathrm{d}}}
\newcommand{\be}{\begin{eqnarray*}}
\newcommand{\ee}{\end{eqnarray*}}

\newcommand{\bee}{\begin{eqnarray}}
\newcommand{\eee}{\end{eqnarray}}
\newcommand{\beeq}{\begin{equation}}
\newcommand{\eeeq}{\end{equation}}